%% file: paper_revised_v2.tex
\documentclass[10pt,aps,prc,floatfix,twocolumn,nofootinbib,superscriptaddress]{revtex4-1}
\usepackage{graphicx,amsmath,amssymb,bm}
\usepackage{amsfonts}

\usepackage[utf8]{inputenc}
\usepackage{verbatim}
\usepackage{float}
\usepackage[labelfont={small},font=small,subrefformat=parens,caption=false]{subfig}
\usepackage{cancel}
\usepackage{multirow}
\usepackage{array}
\usepackage{xparse}
\usepackage{xspace}
\usepackage{bigstrut}

\usepackage{physics}
\usepackage{xcolor}
\usepackage{soulpos}  %\usepackage{soul}
\usepackage{dsfont}
\usepackage{dcolumn}
\newcolumntype{d}[1]{D{.}{.}{#1}}
\usepackage{xurl}
\usepackage[pdfencoding=auto, pdfpagelabels]{hyperref}

\usepackage[overload]{textcase}

\definecolor{linkcolor}{rgb}{0,0,0.40} 
\hypersetup{%
    pdfsubject=Paper,
    pdfkeywords={nuclear physics} {Bayesian} {chiral EFT} {perturbative QCD},
    unicode = true,
    breaklinks = true,
    colorlinks = true,
    linkcolor = linkcolor,
    citecolor = linkcolor,
    menucolor = linkcolor,
    urlcolor = linkcolor
}

\usepackage{cellspace}
\graphicspath{{./paper_figures/}}

\input{buqeye_macros}  % Consistency with all files

\newcommand{\ChiEFT}{$\chi$EFT}

\newcommand{\Lambdabar}{\bar{\Lambda}}

\definecolor{lightblue}{rgb}{.87,.95,.99}
\newcommand{\orcid}[1]{\href{https://orcid.org/#1}{\includegraphics[scale=0.055]{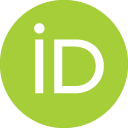}}}

\begin{document}

\title{From chiral EFT to perturbative QCD: a Bayesian model mixing approach to symmetric nuclear matter}

\author{A.~C. Semposki
\orcid{0000-0003-2354-1523}}
\email{as727414@ohio.edu}
\affiliation{Department of Physics and Astronomy and Institute of Nuclear and Particle Physics, Ohio University, Athens, OH 45701, USA}

\author{C. Drischler
\orcid{0000-0003-1534-6285}}
\email{drischler@ohio.edu}
\affiliation{Department of Physics and Astronomy and Institute of Nuclear and Particle Physics, Ohio University, Athens, OH 45701, USA}
\affiliation{Facility for Rare Isotope Beams, Michigan State University, East Lansing, MI 48824, USA}

\author{R.~J. Furnstahl \orcid{0000-0002-3483-333X}}
\email{furnstahl.1@osu.edu}
\affiliation{Department of Physics, The Ohio State University, Columbus, OH 43210, USA}

\author{J.~A. Melendez \orcid{0000-0003-1359-1594}}
\email{melendez.27@osu.edu}
\affiliation{Department of Physics, The Ohio State University, Columbus, OH 43210, USA}

\author{D.~R. Phillips
\orcid{0000-0003-1596-9087}}
\email{phillid1@ohio.edu}
\affiliation{Department of Physics and Astronomy and Institute of Nuclear and Particle Physics, Ohio University, Athens, OH 45701, USA}
\affiliation{Department of Physics, Chalmers University of Technology, SE-41296 G\"oteborg, Sweden}

\date{\today}

\begin{abstract}
Constraining the equation of state (EOS) of strongly interacting, dense matter is the focus of intense experimental, observational, and theoretical effort. Chiral effective field theory ($\chi$EFT) can describe the EOS between the typical densities of nuclei and those in the outer cores of neutron stars while perturbative QCD (pQCD) can be applied to properties of deconfined quark matter, both with quantified theoretical uncertainties. However, describing the full range of densities in between with a single EOS that has well-quantified uncertainties is a challenging problem. Bayesian multi-model inference from $\chi$EFT and pQCD can help bridge the gap between the two theories. In this work, we introduce a correlated Bayesian model mixing framework that uses a Gaussian Process (GP) to assimilate different information into a single QCD EOS for symmetric nuclear matter. The present implementation uses a stationary GP to infer this mixed EOS solely from the EOSs of $\chi$EFT and pQCD while accounting for the truncation errors of each theory. The GP is trained on the pressure as a function of number density in the low- and high-density regions where $\chi$EFT and pQCD are, respectively, valid. We impose priors on the GP kernel hyperparameters to suppress unphysical correlations between these regimes. This, together with the assumption of stationarity, results in smooth $\chi$EFT-to-pQCD curves for both the pressure and the speed of sound. We show that using uncorrelated mixing requires uncontrolled extrapolation of at least one of $\chi$EFT or pQCD into regions where the perturbative series breaks down and leads to an acausal EOS. We also discuss extensions of this framework to non-stationary and less differentiable GP kernels, its future application to neutron-star matter, and the incorporation of additional constraints from nuclear theory, experiment, and multi-messenger astronomy.
\end{abstract}

\maketitle

%%%%%%%%%%%%%%%%%%%%%%%%%%%%%%%%%%%%%%%%%%%%%%%%%%%%%%%%%%%%%%%

\section{Motivation} \label{sec:intro} 

Recent years have seen major observational, theoretical, and experimental improvements in the constraints on the cold dense matter equation of state (EOS). These advances come, e.g., from multi-messenger astronomy~\cite{Vitale:2020rid, Corsi:2024vvr,Koehn:2024set}, nuclear forces derived from chiral effective field theory (\ChiEFT) and their implementation in modern many-body frameworks~\cite{Hammer:2019poc,Tews:2020hgp,Hergert:2020bxy,Drischler:2021kxf}, 
Bayesian quantification of \ChiEFT\ truncation uncertainties~\cite{Melendez:2019izc, Drischler:2020yad, Drischler:2020hwi}, and
novel experimental campaigns such as PREX--II/CREX~\cite{Adhikari:2021phr,CREX:2022kgg} and heavy-ion collisions~\cite{Sorensen:2023zkk,MUSES:2023hyz}.
Computational progress has been facilitated by emulators that can rapidly explore how aspects of \ChiEFT\ nuclear forces affect the properties of nuclear matter~\cite{Melendez:2022kid,Drischler:2022ipa,Jiang:2022oba,Duguet:2023wuh}.
In addition, perturbative Quantum Chromodynamics (pQCD) calculations of cold strongly interacting matter have seen major advances in the past several years: all but one of the several pieces of the $\mathcal{O}(\alpha_s^3)$ contribution to the expansion in the strong coupling constant $\alpha_s$ of the pressure have been computed~\cite{Gorda:2023mkk,Gorda:2022jvk,Gorda:2021kme,Gorda:2018gpy} and the uncertainty due to higher-order terms in the perturbative series quantified~\cite{Gorda:2023usm}. These pQCD results delineate the properties of strongly interacting matter, but only at densities far beyond those of neutron stars.
Indeed, neither \ChiEFT\ nor pQCD gives direct access to the EOS at the densities reached in the inner cores of neutron stars.

Nevertheless, Quantum Chromodynamics (QCD), the theory of strong interactions,
describes strongly interacting matter across all relevant densities.
Bayesian model mixing (BMM) enables statistically principled combination of the predictions for the EOS as computed in the EFT of QCD at nuclear densities, \ChiEFT, and the pQCD EOS. 
In this article, we demonstrate how BMM can be used to construct this combination by applying it to the EOS of isospin symmetric matter at zero temperature. 
More generally, our BMM framework is applicable to the mixing of calculations of the nuclear EOS that have well-quantified uncertainties. It could be applied at different values of the neutron-proton excesses, finite temperatures, etc. As such, this approach can ensure full advantage is taken of the advances listed above, by formulating
global microscopic EOS models covering all densities probed by neutron stars~\cite{NP3M, MUSES:2023hyz}, with fully quantified uncertainties.
Such models are a crucial input to large-scale simulations of supernovae and mergers that shed light on the synthesis of heavy nuclei and the structure and evolution of neutron stars in the universe from first principles~\cite{N3AS}.

BMM is a class of statistical methods that combine outputs from individual models using weights that depend on the location in the input space of the problem~\cite{Coleman, Yao:2021abc, Semposki:2022gcp, yannotty2023model}. 
Here, we generalize the model
mixing procedure described in Refs.~\cite{Phillips:2020dmw,Semposki:2022gcp}, 
%This procedure 
which combines Gaussian random variables that represent the different models---including their error structure---thereby yielding information on the theory underlying both models. In Refs.~\mbox{\cite{Phillips:2020dmw,Semposki:2022gcp}} the combination was made pointwise, but this is problematic in the present application because of the uninformed gap between the mixed models and the neglect of correlations across density in such a pointwise approach.
The approach adopted here
produces a representation of the underlying theory as a set of Gaussian random variables that are \emph{correlated} across the input space, i.e., a Gaussian process (GP). 
GPs are now a standard tool for nonparametric EOS modeling in neutron-star applications~\cite{Landry:2018prl,Essick:2020flb,Mroczek:2023zxo}.
Previous work has emphasized their model-agnostic benefits for constructing EOS prior distributions that are then used for data-driven EOS inference from observational and experimental data. In contrast, here we follow a data-agnostic approach and inform our GP using the low- and high-density limits of QCD,
including discrepancy models, to construct model-driven microscopic EOS posterior distributions.%
\footnote{The terminology here is common in the nuclear astrophysics community but is somewhat inconsistent with the statistics perspective. From the latter perspective, our ``data'' are themselves model predictions with corresponding uncertainties from the low- and high-density limits of QCD, so that some in that community might also call what we are doing in this work ``data-driven EOS inference''.}
This work is an example of BMM in which Gaussian random variables are combined; other versions involve combining means and adjusting the variance to data~\cite{yannotty2023model,Kejzlar:2023tlm}, or forming linear combinations of the probability distribution functions obtained in the individual models~\cite{Coleman,Yao:2021abc}.

Our specific implementation of BMM and its application to symmetric two-flavor strongly interacting matter is described in what follows. Detailed descriptions of the $\chi$EFT and pQCD equations of state at fixed number density $n$ are presented in Sec.~\ref{sec:EOSs}, as is a determination of the truncation uncertainties in both theories. Subsection~\ref{subsec:other_theories} then discusses other theoretical information on the EOS that could be incorporated in our mixing. 
The correlated BMM approach highlighted in this work is described in Sec.~{\ref{sec:curvewise}}, with particular attention given to the role of GP correlation lengths. Constraining these correlation lengths according to expectations regarding the physics of the quark-hadron transition yields larger uncertainties than those obtained in the absence of such constraints.
In Sec.~{\ref{sec:pointwise}}, we illustrate the perils of using a pointwise BMM approach.
We summarize and discuss the implications of our findings for future work in Sec.~\ref{sec:summary}.
Appendix~\ref{ap:KLW} discusses how we obtain the pQCD pressure as a function of the baryon density using the Kohn-Luttinger-Ward formalism~\cite{Kohn:1960zz, Luttinger:1960ua} and Appendix~\ref{ap:cs2_mu} details how we compute the pQCD speed of sound at a given density consistently up to a finite order in $\alpha_s$. 
Our software implementation, including annotated Jupyter notebooks used to produce the results in this paper, is publicly available in a GitHub repository~\cite{EOS_BMM_SNM}.
Throughout the paper, we use natural units in which $\hbar = c = 1$.

%%%%%%%%%%%%%%%%%%%%%%%%%%%%%%%%%%%%%%%%%%%%%%%%%%%%%%%%%%%%%%%

\section{Dense matter equations of state} \label{sec:EOSs}

In this section, we first discuss how we use well-developed machinery for assessing truncation errors of finite perturbation series to quantify each theory's uncertainties  (Subsec.~\ref{subsec:UQ}), and then proceed to the discussion of the microscopic calculations at low ($\chi$EFT in Subsec.~{\ref{subsec:ChiEFT})} and high (pQCD in Subsec.~{\ref{subsec:pQCD})} densities. These calculations become input for multiple-model Bayesian inference (Secs.~\ref{sec:curvewise} and~\ref{sec:pointwise}). Subsection~\ref{subsec:other_theories} discusses other theoretical approaches for the dense-matter EOS whose results could potentially be added to the inference.

\subsection{Uncertainty quantification using the BUQEYE truncation error model} \label{subsec:UQ}

Uncertainty quantification (UQ) is a critical component of microscopic EOS modeling.
In this work, we follow the UQ methodology developed in Ref.~\cite{Melendez:2019izc} and use correlated truncation error estimation to obtain our model uncertainties for each EOS investigated. 

We begin by considering our theories as observables $y(x)$ calculated as a function of the input parameter $x$, which could be the Fermi momentum, the associated density or chemical potential of the matter under consideration. We assume the corresponding $y$ has an expansion in powers of a small parameter $Q(x)$, and then express the cumulative result at the $k$th order in that expansion, $y_k(x)$, as
\begin{equation}
    \label{eq:knownorders}
    y_{k}(x) = y_{\textrm{ref}}(x) \sum_{n=0}^{k} c_{n}(x)Q^{n}(x),
\end{equation}
where $y_{\textrm{ref}}(x)$ is a dimensionful reference scale of the theory that sets the typical size of the observable, and $c_{n}(x)$ are the corresponding coefficients at each order.\footnote{Note that such a perturbative expansion must be valid for the given EOS at least in some region of density or this parameterization will not hold. Also note that $c_1(x) \equiv 0$ in \ChiEFT.} We will make separate choices for $Q(x)$ and $y_{\textrm{ref}}(x)$ for each of the two EOSs considered in this work, 
in each case choosing these aspects of the expansion based on the convergence 
properties and nominal expansion parameters
of the respective theories. 

We desire to quantify truncation uncertainties up to $y_{\infty}$, based on the convergence pattern of the $n \leq k$ orders we possess. To do this, we will need to estimate the impact of orders $n>k$ in our expansion. We  first extract the coefficients $c_{n}(x)$ at each known order, by employing the relations
\begin{eqnarray}
    y_{0}(x) &\equiv& y_{\textrm{ref}}(x) c_0(x), \\
    \Delta y_{n}(x) &\equiv& y_{\textrm{ref}}(x) c_n(x) Q^n(x),         \label{eq:differences}
\end{eqnarray}
where we calculate the correction to $y(x)$ at $n^{th}$ order in Eq.~\eqref{eq:differences}. We can then estimate higher orders based on the lower order coefficients by writing the series expansion for truncated orders as
\begin{equation}
    \label{eq:theoryerror}
    \delta y_{k}(x) = y_{\textrm{ref}}(x) \sum_{n=k+1}^{\infty} c_{n}(x)Q^{n}(x).
\end{equation}

We have previously defined $y_{\textrm{ref}}(x)$ and $Q(x)$, so we only need to find the coefficients in Eq.~\eqref{eq:theoryerror}. Since information on the overall size of the observable and the expansion's convergence is encoded in these quantities, the different coefficients $c_n(x)$ should all be of a similar, $\mathcal{O}(1)$ size. We assume that the $c_n(x)$ curves for $n \leq k$ that we already know via Eq.~\eqref{eq:differences} are independent draws from a Gaussian process (GP). We then make the inductive step that all higher $c_n(x)$ are also drawn from the same GP distribution. 

Gaussian processes can be described as a collection of random variables, any subset of which possesses a joint Gaussian distribution \cite{Melendez:2019izc, rasmussen2006gaussian}. The GPs we use here can be written as 
\begin{equation}
    c_n(x) \sim \mathcal{GP}[0, \bar{c}^{2}r(x, x';\ell)],
\end{equation}
and are composed of a mean function (which we take to be zero) and a positive semi-definite covariance function, $\bar{c}^{2}r(x, x';\ell)$, which is often referred to as the \textit{kernel}.
Kernels are parameterized by a correlation structure $r(x,x';\ell)$ and a marginal variance $\bar{c}^{2}$. These reflect the correlations in the dataset and the variability of the data from the mean, respectively. 
The marginal variance and correlation structure inferred from $\{c_n(x), n \leq k\}$ are then, by assumption, inherited by the uncomputed coefficients that define the missing higher orders in the expansion. For more details on the mathematics of GPs, we refer the reader to Refs.~\cite{rasmussen2006gaussian, Melendez:2019izc}.

For both EOSs discussed in the following sections, the 
squared-exponential radial basis function (RBF) kernel is chosen as the kernel for the truncation-error model. This kernel is described by
\begin{equation}
    r(x, x'; \ell) = \exp\bigg[-\frac{(x - x')^{2}}{2\ell^{2}}\bigg],
\end{equation}
which builds in a strict assumption of smoothness for the $c_n(x)$ across the input space---the RBF kernel is infinitely differentiable. This choice also builds in an assumption of \textit{stationarity} in our input space---the RBF kernel only depends on the relative distance between points in the input space, not on the actual locations of the points themselves.

As a prior on $\bar{c}^2$, we choose the scaled inverse $\chi^{2}$ distribution (see Ref.~\mbox{\cite{Melendez:2019izc}} for more information on this choice). This is a conjugate prior for the variance of the Gaussian distribution, and so is straightforwardly updated using the $c_n(x)$ coefficients
through updating equations for the degrees of freedom $\nu$ and scale $\tau$ of the scaled inverse $\chi^2$. The choice of hyperparameters ($\nu_0$ and $\tau_0^2$), together with other options, are discussed in \cite{Melendez:2019izc}.

The correlations of the $c_n(x)$ across $x$ induces correlations in the truncation error $\delta y_k$ across $x$.\footnote{GPs can also be used to capture correlations across observables, as shown in Refs.~\cite{Drischler:2020hwi, Drischler:2020yad}.}
This desirable feature can be incorporated naturally into the model mixing procedure, where mean predictions $y_k(x)$ and their errors $\delta y_k(x)$ become the training data.
Correlated error structures in this training data stop the mixed model from overweighting information that is actually redundant. 

\subsection{Low-density equation of state from chiral EFT (\texorpdfstring{$\chi$EFT}{ChEFT})} \label{subsec:ChiEFT}

At low densities ($n \lesssim 2n_{0}$), where QCD is highly nonperturbative, \ChiEFT\  provides a systematic expansion for nuclear forces; its expansion parameter is the ratio of the typical momentum of the system and the \ChiEFT\ breakdown scale $\Lambda_b \approx 600 \MeV$ \cite{Epelbaum:2014efa,Drischler:2020yad,Millican:2024yuz}. 
We assert the Fermi momentum as the typical momentum of low-density nuclear matter.

We use here the predictions of \ChiEFT\ up to N$^3$LO for the energy per particle, $E(n)/A$, obtained using fourth-order many-body perturbation theory (MBPT) in Refs.~\cite{Drischler:2017wtt,Leonhardt:2019fua}. 
The (nonlocal) chiral interactions consisted of the family of order-by-order $NN$ potentials developed by Entem, Machleidt, and Nosyk (EMN)~\cite{Entem:2015xwa} and $3N$ forces at the same order in the chiral expansion and with the same momentum cutoff (see Table~I of Ref.~\cite{Drischler:2020yad} for more details). The $3N$ low-energy constants (LECs) $c_{D}$ and $c_{E}$ were fit to the triton binding energy and adjusted to the empirical nuclear saturation point in symmetric nuclear matter~\cite{Drischler:2017wtt}.
However, we stress that the BUQEYE truncation error framework, discussed in Sec.~\ref{subsec:UQ}, can be applied to any order-by-order calculations in the EFT expansion, independent of the computational framework used to solve the many-body Schrödinger equation and the details of the order-by-order EFT interactions.

Throughout this work, we assume both $NN$ and $3N$ LECs to be fixed to the values obtained by the developers of the nuclear potentials.
Specifically, we focus on the pressure from $\chi$EFT up to next-to-next-to-next-to leading order (N$^3$LO; i.e., $k\leqslant 4$), obtained from the
density derivative of $E(n)/A$, for the interactions with the momentum cutoff $\Lambda = 500\MeV$. 
We use these results from Ref.~\cite{Drischler:2020yad}, in which the BUQEYE truncation error model was used to quantify correlated EFT truncation errors and propagate them to derivative quantities such as the pressure. 

\begin{figure}
    \centering
    \includegraphics[width=\columnwidth]{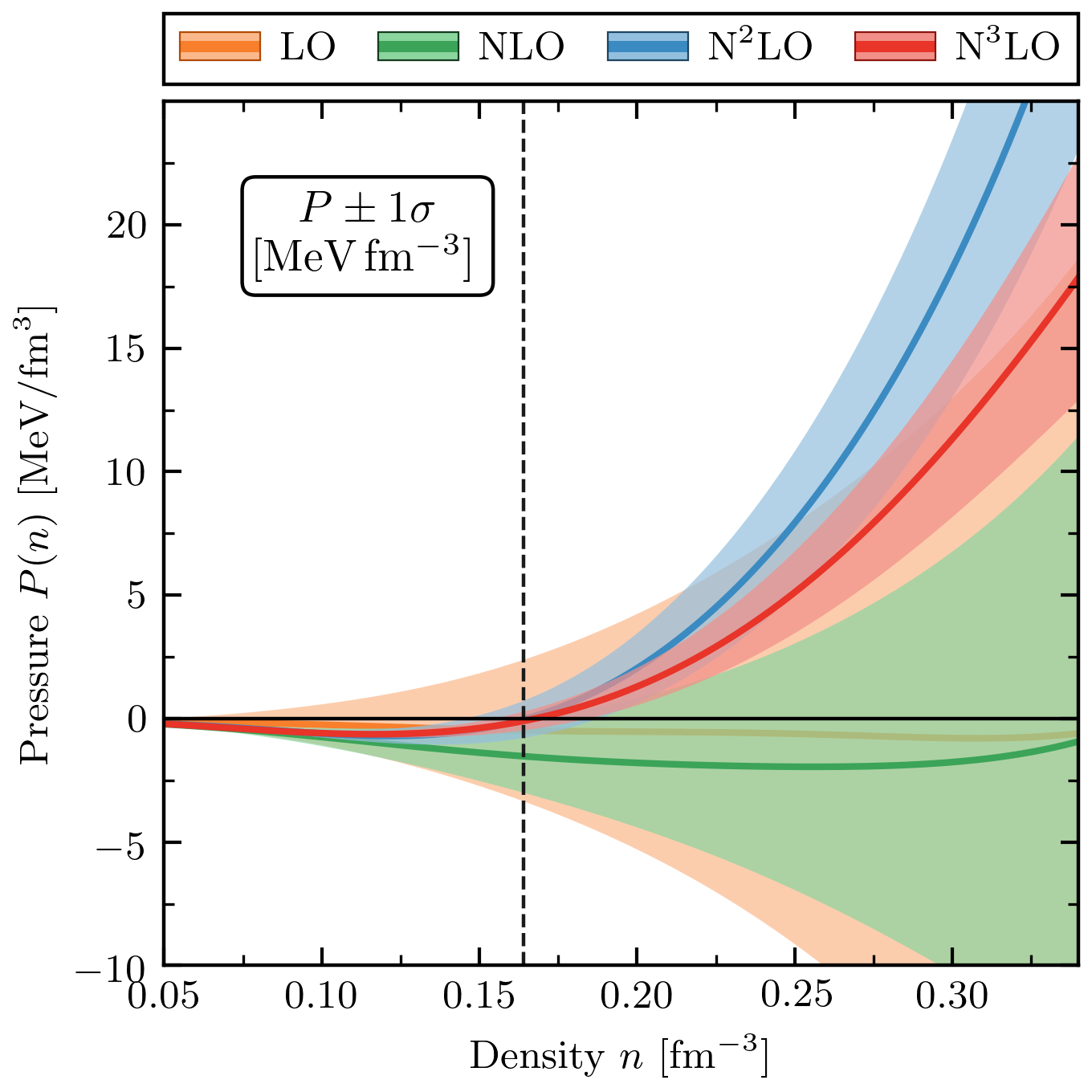}
    \caption{Order-by-order results, including $1\sigma$ uncertainty bands, for the pressure of the isospin-symmetric matter EOS from $\chi$EFT, with an $NN$ potential cutoff of $\Lambda = 500$ MeV. The black dashed line indicates the saturation density, $n_{0} = 0.164$~fm$^{-3}$~\cite{Drischler:2015eba}. Figure code adapted from Jupyter notebooks published with Refs.~\cite{Drischler:2020yad, Drischler:2020hwi}.}
    \label{fig:pressure_n_2n0}
\end{figure}

This was done by first choosing the variable in which the GP would be constructed. For this EOS, this is the Fermi momentum $k_{F}$, as this choice was superior to the density $n$ in terms of stationarity of the coefficients' correlation structure~\cite{Drischler:2020yad}. 
In particular, $y_{\textrm{ref}}$ was chosen to be 
\begin{equation}
    y_{\textrm{ref}}(k_{F}) = 16 \MeV \times \left(\frac{k_{F}}{k_{F, 0}}\right)^{2},
\end{equation}
which was motivated by the structure of the $\chi$EFT leading order (LO) term and empirical knowledge of the nuclear saturation point in SNM. Note here that $k_{F,0} \approx 1.344 ~\textrm{fm}^{-1}$ is the Fermi momentum at the saturation density $n_{0} = n(k_{F,0}) \approx 0.164 \fmiq$~\cite{Drischler:2015eba}, with $n(k_F) = 2 k_F^3/(3\pi^2)$, and  $|E(n_0)/A| \approx 16\MeV$ the corresponding absolute value of the saturation energy in SNM. To employ Eq.~\eqref{eq:knownorders}, an expansion parameter $Q$ must also be specified;
Ref.~\cite{Drischler:2020yad} chose
\begin{equation} \label{eq:Q_def}
    Q(k_{F}) = \frac{k_{F}}{\Lambda_{b}},
\end{equation}
where $\Lambda_b$ is the breakdown scale, estimated to be $\Lambda_b \approx 600 \MeV$ as previously mentioned. 

Results for the pressure of symmetric nuclear matter, $P(n)$, are obtained using
\begin{equation}
    \label{eq:p_chiral}
    P(n) = n^{2} \dv{n} \frac{E}{A}(n),
\end{equation}
see Fig.~\ref{fig:pressure_n_2n0}, which also shows
combined quantified truncation and GP interpolation uncertainties for the prediction of the EFT at each known order. Equation~\eqref{eq:p_chiral} and Fig.~\ref{fig:pressure_n_2n0} are shown in terms of the number density, $n$---we will perform multi-model Bayesian inference in this input space.

To construct the error bands, the scaled inverse $\chi^{2}$ priors on $\bar{c}$ and $\ell$ discussed in the previous section were used, with degrees of freedom $\nu_0 = 10$ and scale parameter $\tau_0^{2} = (\nu_0 - 2)/\nu_0$ to enforce the condition of naturalness of the known coefficients $c_{n}(x)$. A white noise kernel was added to the squared-exponential RBF kernel to account for the higher-order uncertainty from the MBPT calculations (see Ref.~\cite{Drischler:2020yad} for more details). The 68\% (i.e., $1\sigma$) intervals in Fig.~\ref{fig:pressure_n_2n0} indicate good convergence properties at each order around and below $n_{0}$. We point out that though the N$^3$LO band may appear wider than the next-to-next-to leading order (N$^2$LO) band at densities $\simeq 2n_{0}$, this is not the case when comparing the size of these uncertainty bands numerically.

\begin{figure}
    \centering    \includegraphics[width=\columnwidth]{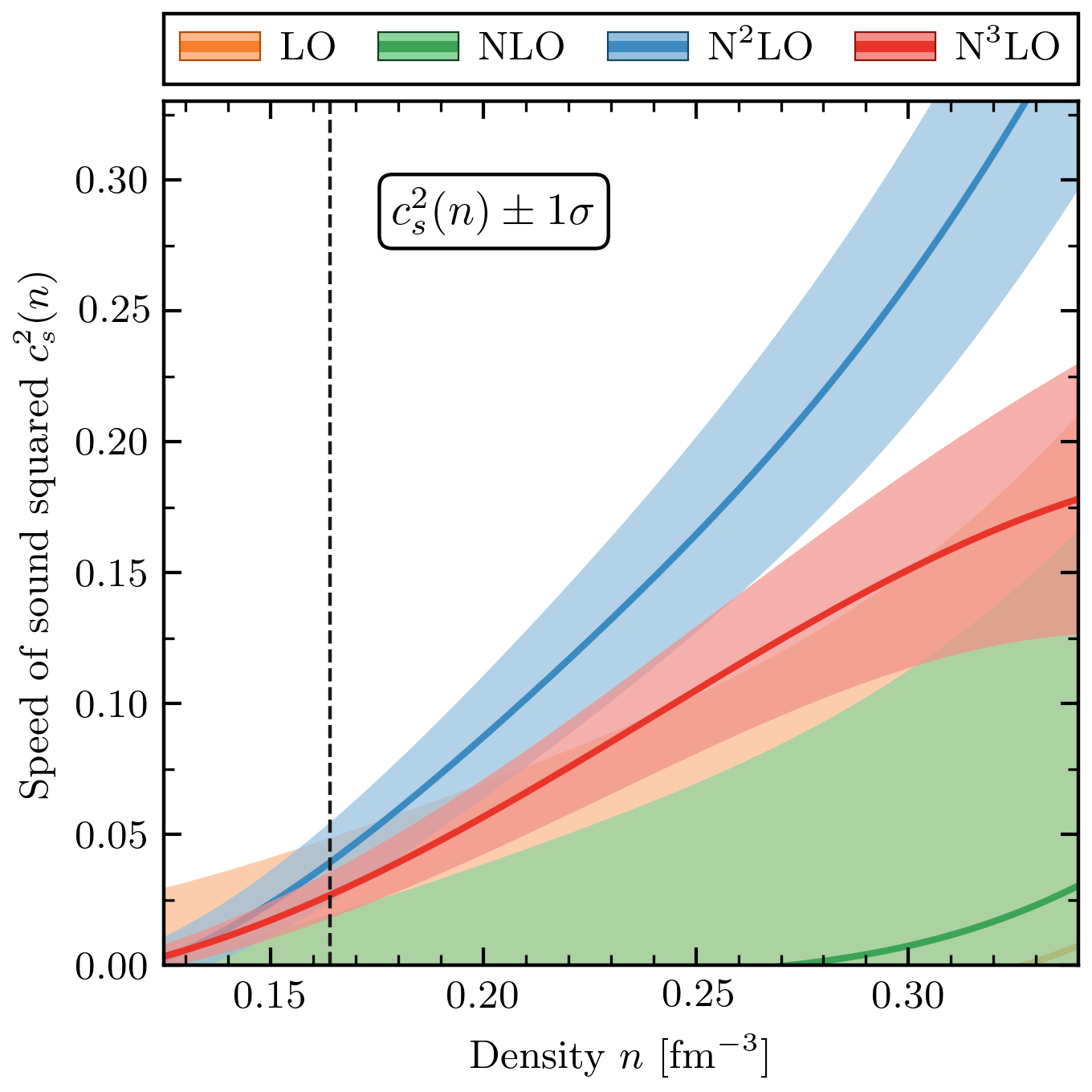}
    \caption{Speed of sound squared results for symmetric matter from $\chi$EFT, sampling directly from $E(n)/A$. The black dashed line indicates the saturation density of $n_{0} = 0.164$ fm$^{-3}$. We only plot above where $c_{s}^{2}(n)=0$, as a negative sound speed indicates unstable matter. Figure code adapted from Jupyter notebooks published with Refs.~\cite{Drischler:2020yad, Drischler:2020hwi}.}
    \label{fig:cs2_n_2n0}
\end{figure}

With $P(n)$ and $E(n)/A$ in hand, we compute the speed of sound from
\begin{equation}
    c_s^2(n)=\frac{\partial P}{\partial \varepsilon}=\frac{\partial P}{\partial n}\left(\frac{\partial \varepsilon}{\partial n}\right)^{-1},
    \label{eq:cs2ofn}
\end{equation}
where $\varepsilon(n) = n \left[E(n)/A + m_N\right]$ is the energy density with the nucleon rest-mass ($m_N$) contribution.
To properly propagate the correlated uncertainties to the sound speed, we sample curves from $E(n)/A$ and compute derivatives that preserve the aforementioned correlations. Figure~\ref{fig:cs2_n_2n0} shows the corresponding results order-by-order through N$^{3}$LO.

\subsection{High-density equation of state from perturbative QCD (pQCD)} \label{subsec:pQCD}

At very high densities ($n \gg 2n_0$),
QCD becomes perturbative due to asymptotic freedom, and thus can be expanded in the strong coupling constant, $\alpha_{s}$. We take this, at the two-loop level, to be \cite{Kurkela:2009gj, ParticleDataGroup:2020ssz}
\begin{equation}
    \label{eq:alpha_s}
    \alpha_{s}(\bar{\Lambda}) = \frac{4 \pi}{\beta_{0} L}\left[ 1 - \frac{2\beta_{1}}{\beta_{0}^{2}}\frac{\ln{L}}{L}\right],
\end{equation}
with 
\begin{eqnarray}
    L &=& \ln({\bar{\Lambda}^{2}/\Lambda_{\overline{MS}}^{2}}), \quad \bar{\Lambda} = 2X\mu, \nonumber \\
    \beta_{0} &=& 11 - \frac{2}{3}N_{f}, \quad \beta_{1} = 51 - \frac{19}{3}N_{f},
\end{eqnarray}
where 
$\bar{\Lambda}$ is the renormalization scale, defined in this work to vary around the chemical potential by a factor $2X = [1, 4]$. For the purposes of this proof-of-principle study we invoke the large-$N_f$ and phenomenological model arguments of Ref.~\cite{Kurkela:2009gj} and set $X = 1$ in the main results of our study.
$\Lambda_{\overline{MS}}$ has been determined via experimental constraints to be $0.38 
\pm 0.03$~GeV~\cite{Kurkela:2009gj}. We note that, because we use this value of $\Lambda_{\overline{MS}}$ at each order of the pQCD result, we must also incorporate the two-loop running of $\alpha_{s}(\Lambdabar)$ throughout the calculation.
    
The form of pQCD EOS used here is the weak-coupling expansion of the theory through N$^{2}$LO.
The parameterization we choose to employ is taken from Ref~\cite{Gorda:2023mkk}:
\begin{eqnarray}
    \label{eq:pressurepqcd2023}
    \frac{P(\mu)}{P_{FG}(\mu)} \simeq &1& +~a_{1,1} \left(\frac{\alpha_{s}(\Lambdabar)}{\pi}\right) \nonumber \\ &+& N_{f}\left(\frac{\alpha_{s}(\Lambdabar)}{\pi}\right)^{2} \bigg[ a_{2,1} \ln\left(\frac{N_{f}\alpha_{s}(\Lambdabar)}{\pi}\right) \nonumber \\ 
    &+& a_{2,2}\ln\frac{\bar{\Lambda}}{2\mu} + a_{2,3}\bigg] + \mathcal{O}(\alpha_{s}^{3}),
\end{eqnarray}
with coefficients $a_{1,1} = -2$, $a_{2,1} = -1$, $a_{2,2} = -4.8333$, $a_{2,3} = -8.0021$, and where $N_{f} = 2$ and $N_{c} = 3$ for the present system.\footnote{For the explicit formulae regarding the pQCD EOS for arbitrary quark colour and flavour, see Eq.~(37) and Table I in the Supplemental Material of Ref.~\cite{Gorda:2023mkk}.}
Here, we also define the Fermi gas (FG) pressure as
\begin{equation}
    P_{FG}(\mu) \equiv N_{c} N_{f}\frac{\mu^{4}}{12 \pi^{2}} = \frac{\mu^{4}}{2 \pi^{2}}.
\end{equation}
In this work, we use $\mu \equiv \mu_{q}$, and denote $\mu_{B} = 3 \mu$ explicitly when discussing the baryon chemical potential.
\begin{figure}
    \centering
     \includegraphics[width=\columnwidth]{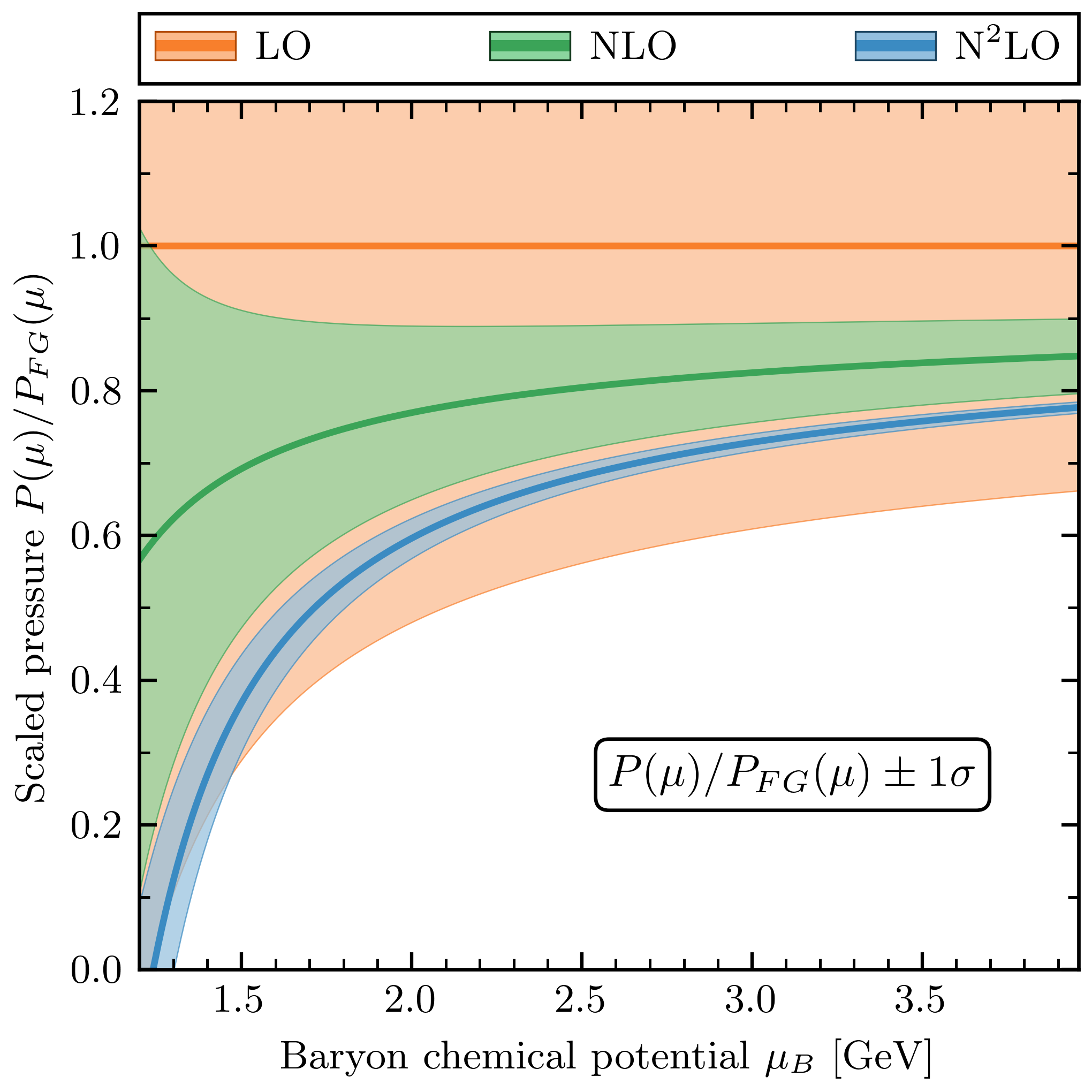}
    \caption{Order-by-order $P(\mu)$ results, scaled with the Fermi gas pressure $P_{FG}(\mu)$, for the EOS from pQCD, plotted against the baryon chemical potential $\mu_{B} = 3 \mu$, where $\mu \equiv \mu_{q}$. Results are shown up to and including LO, NLO, and N$^2$LO, respectively, with $1\sigma$ truncation error bands.}
    \label{fig:pqcd_pressure_mu}
\end{figure}

To properly quantify the truncation error in the pQCD pressure $P(\mu)$, we apply the methodology of Sec.~\ref{subsec:UQ} to the results of Eq.~\eqref{eq:pressurepqcd2023}. We choose an expansion parameter of $Q(\bar{\Lambda}) = N_{f} \alpha_s(\bar{\Lambda})/\pi$, and the reference scale $y_{\textrm{ref}} = P_{FG}(\mu)$, and train on the known coefficients of each order, treating them as functions of the quark chemical potential. This produces the truncation error bands shown in Fig.~\ref{fig:pqcd_pressure_mu} at the $1\sigma$ level. The NLO result is well within the 68\% credibility interval of the LO (Fermi gas) result, but the N$^2$LO $P(\mu)$ curve sits between 1 and 1.5$\sigma$ away from the NLO pressure. We will take this opportunity to emphasize the point that this is not a statistically inconsistent result. Indeed, if every order fell within the previous order's 68\% credibility interval, then that would mean the expansion parameter adopted for the perturbative series was too large, and the resulting error bands too large (``conservative"). 

To perform model mixing with both models in the same input space, we choose to convert the pQCD pressure, $P(\mu)$, to $P(n)$. To do this in a way that retains only terms up to $\mathcal{O}(\alpha_{s}^{2})$, we invoke the procedure of Kohn, Luttinger, and Ward~\cite{Kohn:1960zz, Luttinger:1960ua}.
This procedure first inverts the expression for $n(\mu)$ consistently up to the desired order in perturbation theory, producing a series  $\mu=\mu_{FG} + \mu_1 + \mu_2 + \ldots$ in which each $\mu_k$ equals $\mu_{FG} \alpha_s^k$ times a numerical coefficient. This series is then substituted into the formula for $P(\mu)$ and the result expanded again in powers of $\alpha_s$ and truncated at the desired order. 
This yields a perturbative series for $P$ in which the only chemical potential that appears is $\mu_{FG}$. Since
\begin{equation}
\mu_{FG}=\left(\frac{3 \pi^2 n}{N_c N_f}\right)^{1/3},
\label{eq:muFG}
\end{equation} 
this means we have accomplished our goal of finding the strictly perturbative expression for the function $P(n)$. This calculation is carried out in detail in Appendix~\ref{ap:KLW}, and up to and including $\mathcal{O}(\alpha_{s}^{2})$, it gives
\begin{eqnarray}
    \label{eq:p_nklw}
    \frac{P(n)}{P_{FG}(n)} &=& 1 + \frac{2}{3\pi}\alpha_{s}(\Lambdabar_{FG}) \nonumber \\
    &+& \frac{8}{9\pi^{2}}\alpha_{s}^{2}(\Lambdabar_{FG}) - \frac{N_{f}^{2}}{3\pi^{2}}c_{2}(\mu_{FG})\alpha_{s}^{2}(\Lambdabar_{FG})\nonumber \\
    &-& \frac{\beta_{0}}{3\pi^{2}}\alpha_{s}^{2}(\Lambdabar_{FG}),
\end{eqnarray}
where 
\begin{align}
        c_{2}(\mu_{FG}) = \frac{1}{N_{f}}\bigg[a_{2,1}&\ln{\left(\frac{N_{f}\alpha_{s}(\Lambdabar_{FG})}{\pi}\right)} \nonumber \\ 
        &+ a_{2,2}\ln{\left(\frac{\Lambdabar_{FG}}{2\mu_{FG}}\right)} + a_{2,3}\bigg].
\end{align}
Comparing Eqs.~(\ref{eq:p_nklw}) and~(\ref{eq:pressurepqcd2023}) shows that the NLO contribution to $P(n)$ has the opposite sign to that in $P(\mu)$. Consequently, the direction of approach to the asymptotic Fermi gas result in $P(n)$ is from above, rather than from below, as it is in $P(\mu)$. The NLO coefficient in $P(n)$ is also only 1/3 the size of the corresponding piece of $P(\mu)$, and the N$^2$LO coefficient is also markedly smaller in $P(n)$. These changes in the properties of the expansion are all consequences of performing the KLW inversion from pressure at fixed chemical potential to obtain pressure at fixed number density.

\begin{figure}
    \centering
    \includegraphics[width=\columnwidth]{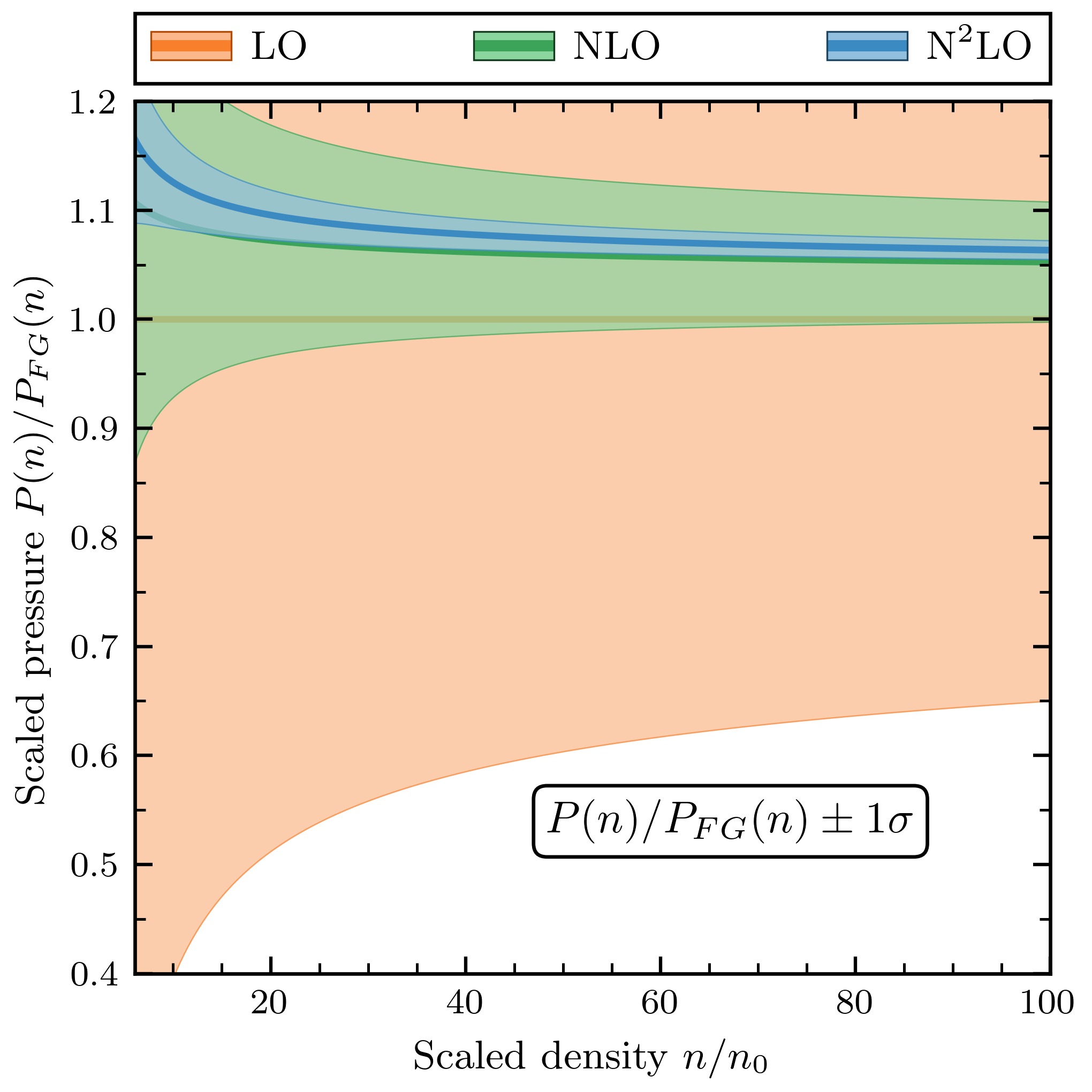}
    \caption{Order-by-order results for the pressure as a function of scaled number density $n/n_{0}$, via the KLW inversion, from the pQCD EOS.}
    \label{fig:pqcd_pressure_n}
\end{figure}

Figure~\ref{fig:pqcd_pressure_n} shows our results for $P(n)$, as well as the application of the BUQEYE truncation error framework discussed in Sec.~\ref{subsec:UQ}. These truncation error bands are obtained by using the same expansion parameter, variance (and length scale) as in the $P(\mu)$ result, but evaluated at the quark chemical potential $\mu_{FG}$ corresponding to each desired baryon number density, Eq.~(\ref{eq:muFG}). Because the coefficients in the expansion for $P(\mu)$ are larger than those in $P(n)$ the credibility intervals obtained by this procedure have empirical coverage probabilities that are higher than their theoretical credibility percentage. 

Recent Bayesian treatments of pQCD uncertainty in the EOS have computed the truncation error (sometimes called the ``missing higher-order uncertainty'') using an approach similar to the one employed here and then marginalized over the unphysical parameter $X$ to obtain a final pdf for the pQCD pressure~\mbox{\cite{Gorda:2023usm, Gorda:2023mkk}}. Because the central value of $P(n)$ at a particular order in pQCD is $X$-dependent, this procedure produces pQCD error bars that are markedly larger than the ones displayed in Fig.~{\ref{fig:pqcd_pressure_n}}, where we made the choice $X=1$.%
\footnote{The size of the pQCD uncertainty is also contingent on the choice of the error model, i.e., the statistical assumptions made regarding the distribution of higher-order coefficients in the series.}
Observables would be independent of the choice of $X$ if the pQCD expansion were carried out to all orders, so this scale dependence is a complementary way to access the truncation error of the pQCD series. Including both scale variation and the missing higher-order uncertainty in the pQCD error budget runs the risk of double-counting uncertainties. References~\mbox{\cite{Cacciari:2011ze,NNPDF:2024dpb}} provide one example of pQCD uncertainty assessment from the series truncation error and one from scale variation. The former work even compares the two approaches. Neither combine them.

\begin{figure}
    \centering
    \includegraphics[width=\columnwidth]{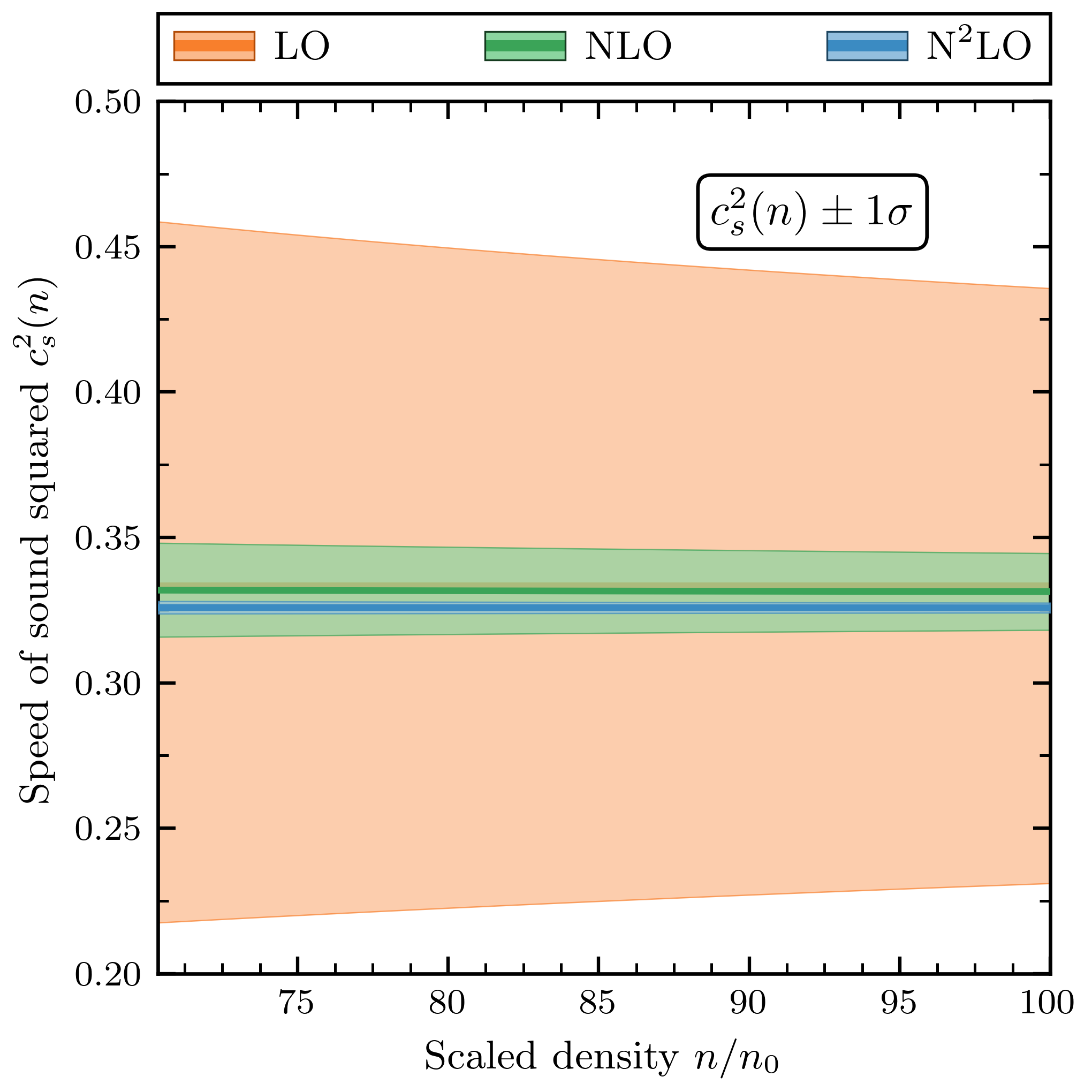}
    \caption{Order-by-order speed of sound squared results for the pQCD EOS with respect to the scaled number density $n/n_{0}$. As discussed in the main text, the result for the Fermi gas (leading order) contribution is exactly 1/3 (orange line), as expected, and the next-to-leading-order contribution (green line) is consistent with zero.}
    \label{fig:pqcd_cs2n.pdf}
\end{figure}

Since, in the context of pQCD, we have $P$ as a function of chemical potential, we can compute the sound speed squared from Eq.~(\ref{eq:pressurepqcd2023}), employing
\begin{equation}
    \label{eq:soundspeedmu}
    c_{s}^{2}(\mu) =  \frac{\partial P}{\partial \mu} \left( \mu~\frac{\partial^{2} P}{\partial \mu^{2}} \right)^{-1}.
\end{equation}
Details of this calculation can be found in  Appendix~\ref{ap:cs2_mu}. Meanwhile, Eq.~(\ref{eq:cs2ofn}) can be recast as 
\begin{equation}
\label{eq:cs2doublelogdvtv}
c_s^2(n)=n \frac{\partial \ln(\mu(n))}{\partial n},
\end{equation}
through the use of $P=n \mu - \varepsilon$ and $\mu=\frac{\partial \varepsilon}{\partial n}$.
In the second part of Appendix~\ref{ap:cs2_mu}, we 
use this compact form to compute $c_s^2(n)$ and find that, up to second order, it has the same functional form as $c_s^2(\mu)$. Moreover, we show that $c_s^2$ approaches its asymptotic value of $1/3$ from below, and, at large density, the deviation of this quantity from 1/3 is of $\mathcal{O}(\alpha_s^2)$, with a coefficient determined purely by the running of $\alpha_s$. 
The $\mathcal{O}(\alpha_s)$ contribution to $c_s^2$ is zero. Figure \ref{fig:pqcd_cs2n.pdf} plots these results for the speed of sound squared as a function of the density, together with their uncertainties. The N$^2$LO result sits slightly below $1/3$ in the range shown here, as expected. We note that in Fig.~\ref{fig:pqcd_cs2n.pdf} the NLO result deviates slightly from 1/3 because the speed of sound is computed using Eq.~(\ref{eq:cs2doublelogdvtv}) here, and the derivatives of pressure in the numerical implementation of that equation induce effects of $\mathcal{O}(\alpha_{s}^2)$ and above in the final result. 

\begin{table}[t]
\renewcommand{\arraystretch}{1.2}
    \setlength{\tabcolsep}{15pt}
    \caption{A selection of common cutoff densities in $\chi$EFT (N$^{3}$LO) at which the relative error of the pressure $P(n)$ of pQCD (N$^{2}$LO) is equal. We evaluate these relative errors at $X=1$, our nominal choice throughout this paper.}
    \centering
    \begin{ruledtabular}
    \begin{tabular}{c|c|c|c} 
     \multirow{2}{*}{$Q(\Lambdabar)$} & \multicolumn{2}{c|}{$n/n_{0}$} & Relative \\ \cline{2-3}
      & $\chi$EFT & pQCD & error (\%) \\
     \hline
     \multirow{3}{*}{$\frac{N_{f}}{\pi}\alpha_{s}(\Lambdabar)$} & 1.5 & 2.43 & 33 \\
      & 1.75 & 2.59 & 29 \\ 
      & 2.0 & 2.62 & 28 
    \end{tabular}
    \end{ruledtabular}
    \label{tab:errorbandcomp}
\end{table}

Table~\ref{tab:errorbandcomp} compares the size of the truncation error bands in $\chi$EFT at $n=1.5 n_0$, $n=1.75 n_0$, and $n=2 n_0$ to the size of the pQCD error bands to determine where in density pQCD becomes as uncertain as \ChiEFT\ is at each of those three densities. (A similar exercise  was carried out in Ref.~\cite{Fraga:2013qra}.)  For the choice $X=1$, and taking an expansion parameter $Q(\Lambdabar) = \frac{N_{f}}{\pi}\alpha_{s}(\Lambdabar)$, this happens $0.5$--$1 n_0$ above the \ChiEFT\ density we choose for the comparison---and therefore much lower than the $(20-40)n_{0}$ that is often quoted (see, e.g., Ref.~\cite{Gorda:2023nvq}). This might lead one to believe that pQCD is applicable at neutron star densities. However, the perturbative series does not include all nonperturbative effects; for example, hadronization and pairing are not accounted for there. Once such effects become sizable, the uncertainty computed by analyzing the order-by-order behavior of pQCD will no longer encompass the full QCD result~\cite{Komoltsev:2023zor, Gorda:2023usm}.

\subsection{Other theoretical approaches to the EOS}

\label{subsec:other_theories}

There are, of course, several models of QCD that provide information on the EOS in the intermediate regime between $2 n_0$ and $40 n_0$. In principle, these can be added to our multi-model inference: the framework is easily extended to inference from more than two models. However, a model can only be added to the mix if it comes with reliable uncertainties---that preferably also have a well-understood correlation structure. In this subsection, we list QCD-based approaches to the EOS that may satisfy this condition, but which we have not yet included in our multi-model inference.  

\begin{itemize}
    \item \textit{Constraining the EOS at low densities using pQCD.} The requirements that strongly interacting matter have a thermodynamically stable and causal EOS have been used, together with the pQCD EOS at densities above $40 n_0$, to construct likelihoods in the $\varepsilon-P$ plane for the strongly interacting EOS~\mbox{\cite{Komoltsev:2021jzg,Komoltsev:2023zor}}. These are results that incorporate additional information beyond the pQCD results for $P(n)$, so we have not included them in our model mixing here.
    
    \item \textit{Lattice QCD constraints.} QCD inequalities imply that the pressure of symmetric nuclear matter at chemical potential $\mu$ is bounded from above by the pressure of fully isospin-stretched matter at $2/3 \mu$~\cite{Moore:2023glb,Fujimoto:2023unl, Abbott:2024vhj}. Reference~\cite{Fujimoto:2023unl} combined this insight with the results for the energy of a system of several thousand pions~\cite{Abbott:2023coj} to place bounds on the symmetric-matter EOS $P(\mu)$. These bounds, when interpreted as a probability distribution, are a $\theta$ function. They do not affect our multi-model inference, i.e., all EOSs we construct satisfy the bound at high credibility levels. 
    
    \item \textit{Functional Renormalization Group (FRG) results.} FRG attempts to truncate the hierarchy of Schwinger-Dyson equations for the $n$-point functions of QCD, by matching to pQCD at a high-resolution scale, and including non-perturbative information on those $n$-point functions at low-resolution scale~\cite{Leonhardt:2019fua, Braun:2020bhy, Braun:2022olp}. Variation of the input choices in the FRG calculation provides a lower bound on its uncertainty. Results for isospin-symmetric QCD matter using this approach---including error bands---were provided in Ref.~\cite{Leonhardt:2019fua}. We display these results on some of our plots, but we emphasize that the error band shown there should not be interpreted as a Bayesian credibility interval. In consequence, it is not clear to us what uncertainties to assign to these FRG calculations, and so we choose not to mix them with the pQCD and \ChiEFT\ results obtained earlier in this section. 
\end{itemize}

%%%%%%%%%%%%%%%%%%%%%%%%%%%%%%%%%%%%%%%%%%%%%%%%%%%%%%%%%%%%%%%%%%
\section{Correlated multivariate model mixing} \label{sec:curvewise}

In this section, we discuss our approach to correlated Bayesian model mixing, first building the formalism in Sec.~\ref{sec:GPformalism}, and then proceeding to our implementation and results in Sec.~\ref{sec:GPresults}.

\subsection{Formalism} \label{sec:GPformalism} 

In the formalism that follows, we express the correlated model mixing procedure through the use of a Gaussian process (GP). This is a prior on the underlying function space for the EOS. The influence of training data on the posterior distribution of the GP is non-local, and it is this aspect of the GP that enables us to predict across the intermediate density region between $\chi$EFT and pQCD.

To define training data for the GP we take advantage of the fact that we have correlated uncertainties for each of the theories that are input to our BMM. Since both theories use GPs to model these uncertainties, we can adopt a GP representation for each theory that holds within that theory's domain of validity.
The pQCD and \ChiEFT\ results can then be expressed as
\begin{equation}
    \label{eq:predictions_curve}
    Y^{(i)}(x) = F(x) + \delta Y^{(i)}(x), \qquad i \in [1, M],
\end{equation}
where $Y^{(i)}(x)$ is a random variable that represents the predictions at a point $x$ from a given model $i$, $F(x)$ is the underlying theory, and $\delta Y^{(i)}(x)$ represents the error of the model. In this work, $i$ denotes \ChiEFT\ or pQCD (hence $M=2$), and $\delta Y^{(i)}(x)$ represents their truncation error, which was modeled as a Gaussian process in Sec.~\ref{subsec:UQ}:
\begin{equation}
\label{eq:GPerrormodel}
    \delta Y^{(i)}(x) \sim \textrm{GP}[0, \kappa_{y}^{(i)}(x,x')],
\end{equation}
where $\kappa_{y}^{(i)}(x,x')$ is the covariance function that defines theory $i$'s correlations between locations $x$ and $x'$ in the input space. 

Equations~(\ref{eq:predictions_curve}) and (\ref{eq:GPerrormodel}) define 
probability density functions (pdfs) for each theory $Y^{(i)}(x)$ that is input to the model mixing: $\pr(y^{(i)}(x) \given f(x),\kappa_y^{(i)})$. We will use Bayes' theorem to combine the information in these pdfs and infer their common mean function $F$.
This requires a prior on the underlying theory, and we also adopt a GP for that:
\begin{equation}
    F(x) \sim \textrm{GP}[0, \kappa_{f}(x,x')],
\end{equation}
although other choices are possible. The model mixing is thus ``non-local" or ``curvewise" because this GP will propagate information across the input space. The reach of the non-locality is determined by the GP hyperparameters, especially the length scale. The hyperparameters, which are  specified in Sec. {~\ref{sec:GPresults}}, are estimated from the training data, and should also be constrained by our understanding of the underlying physics.

The non-local character of the mixing makes it unnecessary to have calculations from both input theories at every point of interest in the input space. Hence, a small number of points can be used to represent each theory, and these can be chosen in the region where the theory is reliable, with the information from that region propagating across the input space through the GP kernel.

In keeping with the GP flavor of these equations, we define a set of training and evaluation data. The training points should be taken to be locations where the input models, and their intra-model UQ, are reliable. The full training data set $\vec{x}_t$ is the concatenation of the individual training data points $\vec{x}_{t,i}$; $i=1, \ldots, M$. In our case, no duplicate entries exist in this vector, since the domains of validity of the input theories are disjoint. We also form the vector $\vec{y} = y(\vec{x}_{t})$ of data in our training space in an analogous manner, i.e., as the concatenation of $y^{(i)}(\vec{x}_{t,i})$; $i=1, \ldots, M$.

Meanwhile, we construct a multi-model theory uncertainty covariance matrix 
\begin{align}
    K_{y} = \kappa_{y}^{(1)}(\vec{x}_{t,1}, \vec{x}_{t,1}) &\otimes \kappa_{y}^{(2)}(\vec{x}_{t,2}, \vec{x}_{t,2}) \otimes \ldots \nonumber \\
    &\otimes \kappa_{y}^{(M)}(\vec{x}_{t,M}, \vec{x}_{t,M})
\end{align} 
that has $M$ block diagonal pieces, each of which 
contains an individual model's truncation error matrix, evaluated at the training points appropriate to that model.\footnote{In this and the following section, we use the convention of indicating random variables with uppercase variables and the values of these random variables with lowercase variables. Note, though, that $K$ is capitalized, not because it is a random variable---it isn't---but because it is a matrix.} Including correlations between the input models is formally a straightforward extension of the assumption that $K_y$ is block diagonal, but here we assume the $M$ models being mixed are independent. However, this does not affect the general formalism, since none of the equations below are dependent on this fact.

The evaluation points are then the locations where we want the mixed-model prediction. We combine the training and evaluation points into a single vector $\vec{x} = \{\vec{x}_{e}, \vec{x}_{t}\}$, and denote the mixed-model predictions at these locations as $\vec{f} = \{\vec{f}_{e}, \vec{f}_{t}\}$. We will carry out the Bayesian inference on the full vector $\vec{f}$ and then marginalize over the values $\vec{f}_t$ at the end of the calculation. We also take $K_f \equiv \kappa_f(\vec{x},\vec{x})$ to be the covariance matrix associated with the function prior, $\kappa_f$, evaluated at the full set of evaluation and training points $\vec{x}$. 
We then define a projection matrix $B_{t}$ that pulls $\vec{f}_{t} = B_{t}\vec{f}$ out of the full vector $\vec{f}$ that includes the evaluation points. The matrix $B_e=1-B_t$ selects the evaluation points from $\vec{f}$.

With these definitions in hand, we can form the log posterior of $\vec{f}$, up to constants, as
\begin{align}
\label{eq:fposterior}
    & \log \pr(\vec{f} \given \vec{y}, K_y, K_f) \nonumber \\ 
    & = -\frac{1}{2}\Bigl[ (\vec{y} - B_t\vec{f}\,)^T K_y^{-1}(\vec{y} - B_t \vec{f}\,) + \vec{f}^T K_f^{-1} \vec{f}\, \Bigr] + \ldots \notag \\
    & = -\frac{1}{2}(\vec{f} - \vec{\mu})^T \Sigma^{-1} (\vec{f} - \vec{\mu})  + \ldots,
\end{align}
so we have shown that $\vec{f}$ is indeed distributed as a multivariate Gaussian:
\begin{align}
  \vec{F} \given \vec{y}, K_y, K_f \sim \normal[\vec{\mu}, \Sigma]
\end{align}
and the algebra that recasts the first line of Eq.~(\ref{eq:fposterior}) in the form of the second line shows that
\begin{align}
   \vec{\mu} & \equiv \Sigma B_t^T K_y^{-1} \vec{y}, \\
   \Sigma & \equiv (K_f^{-1} + B_t^T K_y^{-1} B_t)^{-1}.
\end{align}
Here, we have the inclusion of a full covariance matrix for both $\kappa_{f}$ and each of the models being mixed (via $K_y$). 

We now execute the final step of the calculation, which is to focus on the evaluation points, and exclude the training information from $\mu$ and $\Sigma$. After using the Woodbury matrix identity and applying $B_{e}$ to single out the evaluation points in the above equations, we obtain
\begin{align}
    \label{eq:evaluationdata}
    \vec{\mu}_e & = K_{f,et} (K_{f,tt} + K_{y,tt})^{-1} \vec{y}, \\
    \Sigma_{ee} & = K_{f,ee} - K_{f,et} (K_{f,tt} + K_{y,tt})^{-1} K_{f,te},
\end{align}
where we use the training and evaluation subscripts as before. 
It is no coincidence that these equations have the exact form of the conditional updating equations of a Gaussian process, i.e.,~\cite{rasmussen2006gaussian}
\begin{eqnarray}
    \label{eq:conditionaleqns}
    \vec{\mu}_{x_{*}} &=& K_{f, x_{*}x} (K_{f,xx} + K_{y,xx})^{-1}\vec{y}, \nonumber \\ 
    \Sigma_{x_{*}x_{*}} &=& K_{f,x_{*}x_{*}} - K_{f,x_{*}x} \nonumber \\ &&\null\times (K_{f,xx} + K_{y,xx})^{-1}K_{f,xx_{*}},
\end{eqnarray}
where $x$ are the training points, $x_{*}$ are the evaluation points, and $K_{ij} \equiv \kappa(\vec{x}_{i},\vec{x}_{j})$. 

We have derived the GP conditional updating formulae from our initial Bayesian analysis. It follows that we have obtained a GP representation of the mixed model $F$. That GP is formed from the covariance structure of the input models, evaluated at the training points, and the prior on $F$. These covariance matrices together determine how the models should be mixed.  

\subsection{Implementation and results} \label{sec:GPresults}

To apply the formalism of the previous subsection, we must choose the prior on the underlying theory $\kappa_{f}(x,x')$. Here, we adopt the squared-exponential Radial Basis Function (RBF) kernel with a marginal variance $\cbar^2$. This choice of prior for the function space of $F$ carries with it several assumptions. First, $F$ is assumed to be infinitely differentiable.
%: discontinuous phase transitions are precluded by this aspect of the prior.\footnote{Note that smooth (e.g., hadron-to-quark) crossover transitions are consistent with this approach~\cite{Baym:2017whm}. See, e.g., Refs.~\cite{Kapusta:2021ney,Constantinou:2021hba} for recent work.} 
Second, this is a stationary kernel. An important technical point is that the GP is formulated in $\ln({n})$ space, rather than in the density $n$ itself. This is essential if pQCD is to be described by a stationary GP, since $\alpha_s$ runs logarithmically. Nevertheless, the assumption of stationarity means that we are implicitly assuming there is persistence in both the size and length scale of the variability in the dense matter EOS.
{\it Neither of these assumptions have to be satisfied by QCD.} And indeed, they preclude discontinuous phase transitions. (Smooth, e.g., hadron-to-quark, crossover transitions are consistent with our assumptions~\cite{Baym:2017whm}. See, e.g., Refs.~\cite{Kapusta:2021ney,Constantinou:2021hba} for recent work.)  But, as usual, Bayesian methods provide a clear way to delineate their consequences for the quantity of interest. We are presently investigating the effect that relaxing the assumptions implemented here has on the inference of the dense matter EOS at intermediate densities~\cite{Semposki:2025}. 

We reiterate that because the GP for $F$ can be (actually must be!) evaluated away from its training points, we do not need $\chi$EFT and pQCD to be evaluated at all the points in density for which we wish to know the mixed model result.
We use, in fact, a fairly small training set of four to five points that are clearly in the pQCD domain ($n \geq 20 n_0$) and four points that are clearly in the \ChiEFT\ domain ($n \leq 2 n_0$). The training data is listed in Table~\ref{tab:trainingset}. At least three training points below $2 n_0$ are needed in order to describe \ChiEFT\ well; we also verified that adding more pQCD training points within the range $40 n_0 < n < 100 n_{0}$ does not change the result. As previously discussed in Sec.~\ref{subsec:UQ}, providing the full single-model correlation matrix at the training points chosen from each model ensures the mixing does not double-count correlated information from a particular model. 

We know that nonperturbative effects, not accounted for in the pQCD truncation-error model, will eventually significantly alter the pQCD EOS as we move to lower and lower densities. But, we do not know where these effects reach a magnitude large enough to invalidate our truncation-error model. Because of this, we choose to adopt two ``priors'' regarding the validity of errors derived from pQCD convergence. We train on pQCD predictions~(1) $\geq 20n_0$ and~(2) $\geq 40n_{0}$. We then investigate the sensitivity of the resulting mixed model to this cutoff.

\begin{table}
\renewcommand{\arraystretch}{1.2}
    \caption{The set of training data taken from $\chi$EFT (N$^{3}$LO) and pQCD (N$^{2}$LO) to train the correlated mixing GP. Here we list the location in the scaled density $n/n_{0}$, and the mean and variance at each point. We report values for the two cutoffs used in the pQCD results of this section. We note that though we have only included the variances in the table, we use the full covariance matrices for these points in our training procedure.}
    \centering
    \begin{ruledtabular}
    \begin{tabular}{Sc|d{2.4}|d{2.4}|d{2.4}}
    \multicolumn{1}{Sc}{}& \multicolumn{1}{Sc}{$n/n_{0}$} & \multicolumn{1}{Sc}{$P(n)/P_{FG}(n)$} & \multicolumn{1}{Sc}{$\sigma_{y}(n)$} \\ \hline
    & 0.78 & -0.0257 & 0.007 \\
    \multirow{2}{*}{$\chi$EFT} & 1.14 & 0.0171  &  0.015 \\
    & 1.49 & 0.0830 &  0.028  \\
    & 1.85 & 0.160 &  0.045  \\ \hline
    & 20.11 & 1.095 & 0.023 \\
    \multirow{2}{*}{pQCD} & 36.75 & 1.08 & 0.015 \\
    \multirow{2}{*}{$(20n_0)$} & 53.40 & 1.073 & 0.012 \\
    & 70.04 & 1.068 & 0.010 \\
    & 86.69 & 1.065 & 0.009 \\ \hline
    & 40.08 & 1.078 &  0.014 \\
    pQCD & 56.73 &  1.072 & 0.012 \\
    $(40n_0)$ & 73.37 & 1.068 &  0.010  \\
    & 90.01 & 1.065 &  0.009  \\
    \end{tabular}
    \end{ruledtabular}
    \label{tab:trainingset}
\end{table}

\begin{figure}[t]
    \centering
    \includegraphics[width=\columnwidth]{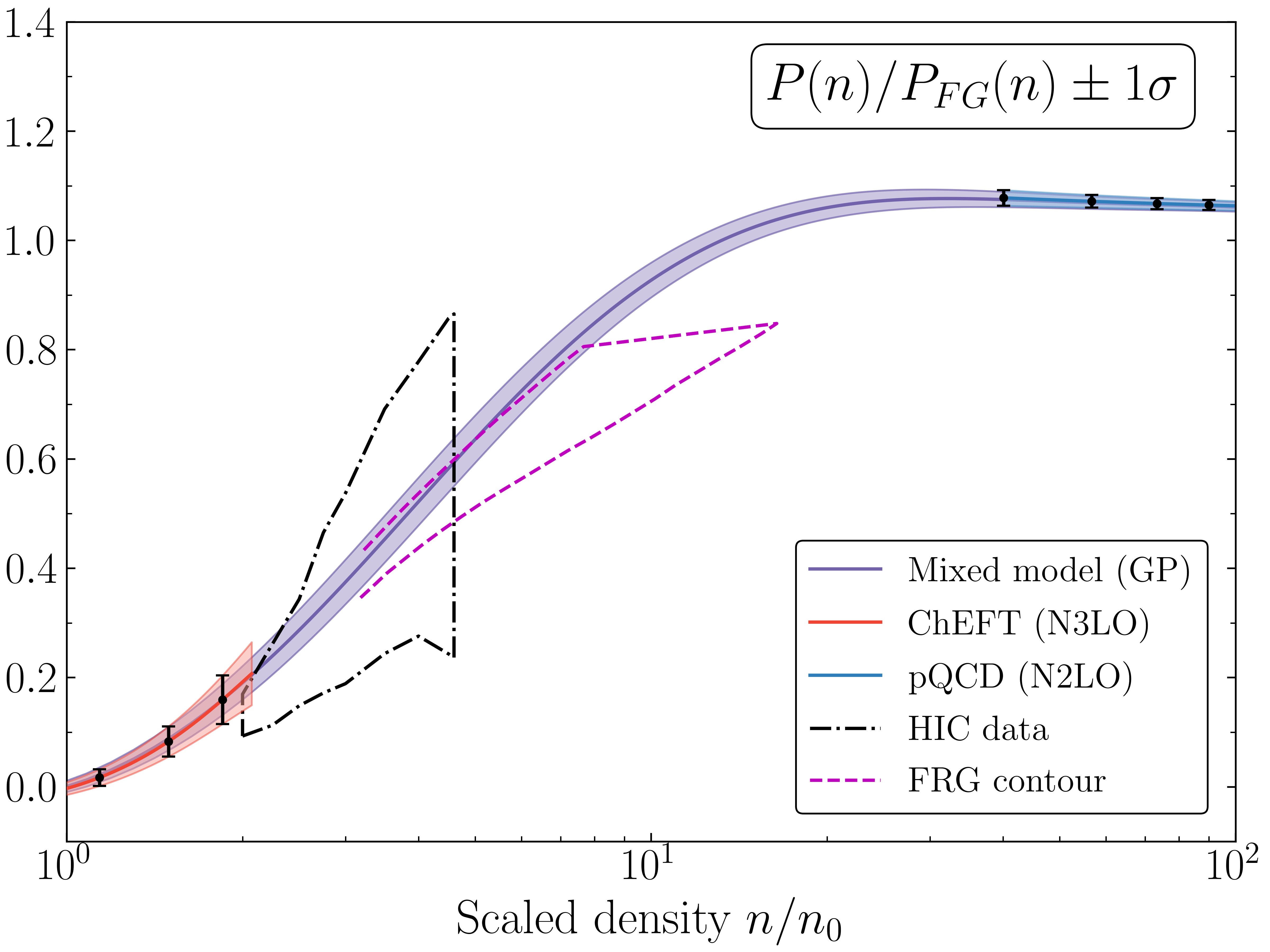}
    \caption{Mixed model results from combining $\chi$EFT and pQCD results for scaled pressure $P(n)/P_{FG}(n)$ using the correlated model mixing approach (Sec.~{\ref{sec:curvewise}}) via a ``naive" Gaussian process, where none of the hyperparameters have any constraints placed upon them. Here we implement pQCD training cutoff of $40n_{0}$. We overlay the results we obtain (purple band) with those of heavy-ion collision data in symmetric nuclear matter (black dash-dot contour) \mbox{\cite{Danielewicz:2002pu}}, and results from the functional renormalization group (FRG) (pink dashed contour) \mbox{\cite{Leonhardt:2019fua}}.} 
    \label{fig:joint_subplot_curvewise}
\end{figure}

\begin{figure*}[t]
    \centering
    \includegraphics[width=\textwidth]{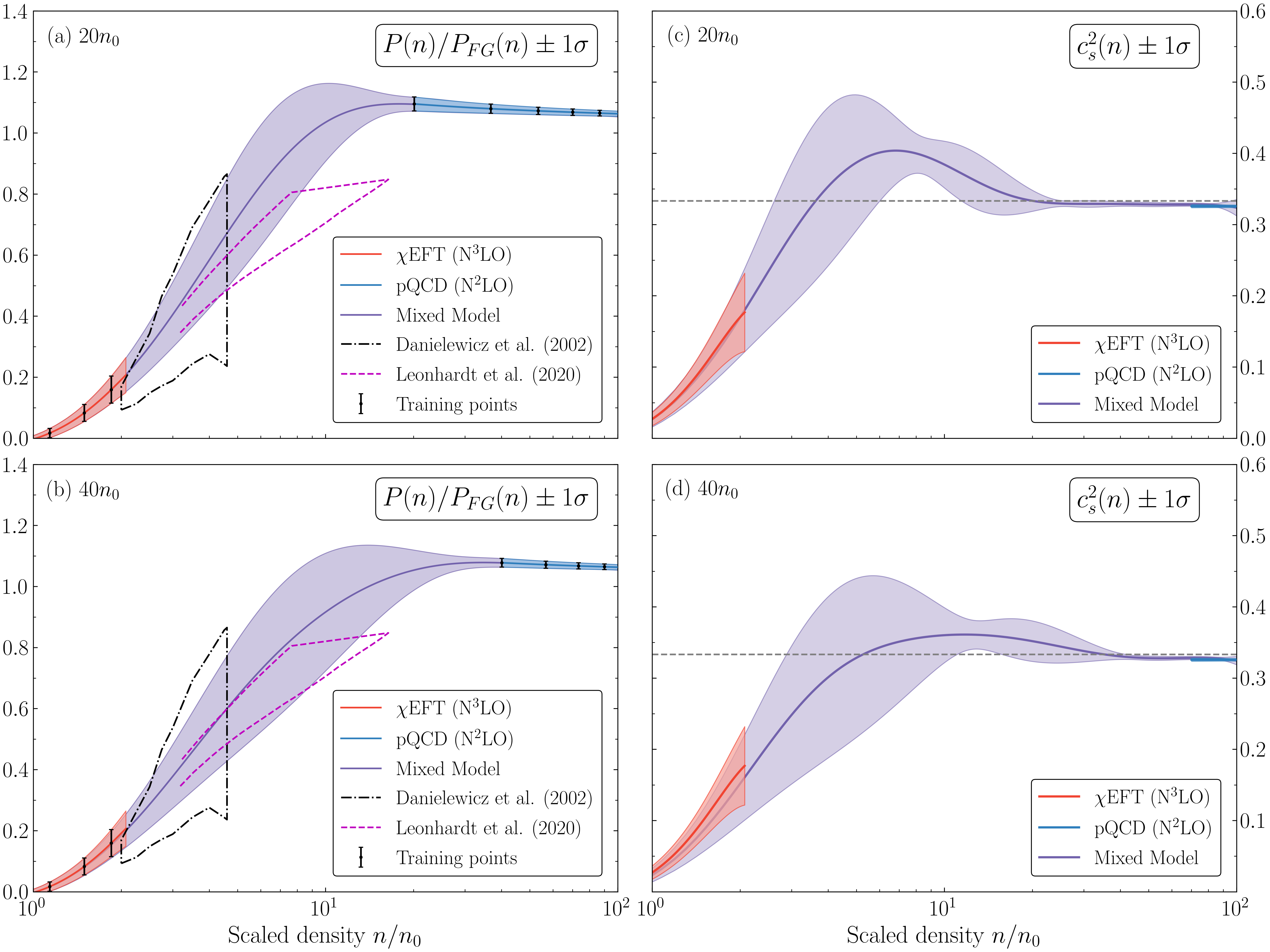}
    \caption{The results for the mixed model EOS using physics-informed hyperpriors on the squared-exponential RBF kernel hyperparameters. (a), (b): the pressure $P(n)$ of the EOS, scaled by the Fermi gas pressure $P_{FG}(n)$. The hyperpriors penalize the unreasonably long correlation lengths of Fig.~{\ref{fig:joint_subplot_curvewise}}, so yielding a much wider uncertainty band, and better reflecting our knowledge about the intermediate region. We note consistency with the heavy ion collision and FRG contours if we use pQCD training data for $n > 20 n_0$ (a) or $n > 40 n_0$ (b); (c), (d): the corresponding speed of sound squared curves as a function of density. The results reproduce the speed of sound (including uncertainties) obtained in pQCD at asymptotically high densities, and overlap quite well the speed-of-sound uncertainty band of $\chi$EFT at low densities.}
    \label{fig:informed_priors}
\end{figure*}

In the following analysis, we will build our mixed model result in two stages: first, we do not incorporate any hyperprior constraints on the length scale or marginal variance of the mixed model GP; then, we apply constraints motivated by the transition between quark and hadron degrees of freedom. In the first stage we implement the GP for $F$ on the scaled $P(n)/P_{FG}(n)$\footnote{This quantity allows for a smoother crossover when model mixing---it eliminates the massive difference in scale between the \ChiEFT\ and pQCD pressures.} data sets using a modified version of the \texttt{scikit-learn} Python package~\cite{scikit-learn}, optimizing the hyperparameters of the prior using the maximum likelihood estimation (MLE) routine already contained in \texttt{scikit-learn}. To train on the data set we have chosen, we create a total covariance matrix out of the two models' respective covariance matrices, such that the resulting matrix input to the GP is block diagonal in model space. We then fit to the training data in Table~\ref{tab:trainingset} in $\ln(n)$ space and predict at a dense grid across the density space, up to $100n_{0}$. Figure~\ref{fig:joint_subplot_curvewise} shows the results from training on \ChiEFT\ and pQCD using the second cutoff previously discussed ($40n_{0}$). The length scale that we found to produce the maximum likelihood for this training data set is: 
$\ell = 1.86$ (in $\ln(n)$). The corresponding marginal variance is 
$\bar{c}^{2} = 0.603$.

The uncertainty on the mixed-model EOS, seen in Fig.{~\ref{fig:joint_subplot_curvewise}}, is rather small. This occurs because the kernel's length scale of 1.86 (in $\ln(n/n_{0})$) results in the training data at, say, $n < 2 n_0$ continuing to have a significant influence on the EOS at densities of order $10 n_0$. Conversely, the mixed-model prediction for the EOS is still markedly correlated with the pQCD training data at densities $\sim 5 n_0$, with a correlation coefficient of 0.355 between $P(5 n_0)$ and $P(40 n_0)$. This persistence of information across a wide density range leads to a marked difference between the error bars of the GP mixed model and those of the individual theories in their respective regions. This is clearest for $P(n)/P_{FG}(n)$ at $\approx 2n_{0}$ in Fig.~{\ref{fig:joint_subplot_curvewise}}: the difference between the mixed-model and \ChiEFT\ bands at this density is approximately 44\%. Such a large impact of pQCD $P(n)$ training data on $P(n)$ in the low-density regime is unrealistic. Naturally, this problem becomes worse if we train on pQCD input down to $n=20 n_0$; we do not show results from that training here.

This problematic correlation structure emerges because we 
determine the hyperparameters of the mixed model GP by maximizing the log-likelihood. To enforce small covariances between the \ChiEFT\ and pQCD regions in the EOS posterior pdf we place a hyperprior (here, a truncated log-normal distribution) on the GP length scale so that 
the distance between the two theories' respective regions is two to three correlation lengths. This choice enforces a transition from nucleon to quark degrees of freedom within the intermediate density range because the region of influence of the training data on the GP posterior is limited by the hyperprior. We also impose a truncated log-normal prior on the marginal variance based on the expectation that it is of natural size.
We then employ the \textit{maximum a posteriori} (MAP) values for correlation length and marginal variance instead of the maximum likelihood ones.

Our subsequent results are shown in Fig.{~\ref{fig:informed_priors}}. The mixed model for $P(n)/P_{FG}(n)$ is in very good agreement with the EOS constraints from HIC data~\cite{Danielewicz:2002pu} for both EOSs (1) and (2) (i.e. both sets of pQCD training data). It falls within the FRG EOS constraint~\cite{Leonhardt:2019fua} in the case where we take training data only at $n \geq 40n_{0}$, and partially for the case of $20n_{0}$. Unsurprisingly, given the kernel choice adopted for $\kappa_f$ here, the mixed model predicts a smooth crossover from $\chi$EFT to pQCD in both cases. The very slight difference in uncertainty between the bands from $\chi$EFT and the mixed model GP is understood since a reduction of uncertainty always occurs when a common mean is inferred from two (or more) independent random variables.

\begin{table*}[t]
    \setlength{\tabcolsep}{29pt}
    \renewcommand{\arraystretch}{1.2}
    \caption{A set of single values of each EOS pictured in Fig.~\ref{fig:informed_priors}. As in the main text, $\chi$EFT is taken at N$^3$LO, and pQCD at N$^2$LO. The mixed model is the correlated Bayesian model mixing result from the GP implementation in Sec.~\ref{sec:GPresults}, calculated using the two chosen lower cutoffs of the pQCD training data (Case (1): $n \geq 20n_{0}$ and Case (2): $n \geq 40n_{0}$). The chemical potential is computed using the perturbative expansion in the KLW formalism.
    Higher-precision, machine-readable data can be found in our GitHub repository~\cite{EOS_BMM_SNM}.}
    \begin{tabular}{c|c|c|c|c}
    \hline\hline
     & & \multicolumn{3}{c}{$P(n)$ [MeV/fm$^3$]} \\ \cline{3-5}
      $n/n_{0}$ &  $\mu_{q}$ [GeV] & pQCD & \multicolumn{2}{c}{GP} \\ \cline{3-5}
      & & $X=1$ & Case (1) & Case (2) \\ \hline
    4 & 0.554 & 253.8 (30.7) & 120.9 (32.7) &  109.6 (31.5) \\
    12 & 0.696 & 1000.1 (31.9) & 959.9 (76.8) & 863.8 (152.3) \\
    36 & 0.958 & 4188.9 (59.3) & 4186.0 (59.2) & 4182.7 (59.5) \\
    \hline\hline
    \end{tabular}
    \label{tab:onepointeos}
\end{table*}

To provide easy access to our results, in Table~\ref{tab:onepointeos}, we provide the mean and standard deviation of the $P(n)$ distribution in our mixed-model EOS at densities of $4n_{0}$, $12n_{0}$, and $36n_{0}$. We list values for the EOS trained on both pQCD training cutoffs ($\geq 20n_{0}$ and $\geq 40n_{0}$). For comparison, we also provide the pQCD $P(n)$, and its uncertainty, at these same densities for our canonical choice of $X=1$. Full results are found in the open-source GitHub repository that accompanies this work~\cite{EOS_BMM_SNM}.

To obtain the speed of sound squared, we sample curves from the mixed result for $P(n)/P_{FG}(n)$, and integrate the corresponding $P(n)$ curves to obtain the energy density, $\varepsilon(n)$, according to:
\begin{equation}
        \label{eq:energydensityintegral}
        \varepsilon(n) = n \left(\frac{\varepsilon(n_{i})}{n_{i}} + \int_{n_{i}}^{n} \frac{P(n')}{n'^{2}} \dd{n'} \right),
    \end{equation}
where $\varepsilon(n_{i})$ is the value of the energy density at a particular point $n_{i}$ in the density space. We set $\varepsilon(n_{i})$ to the value of the energy density at $n_{i} = 100n_{0}$ in the pQCD EOS, as this yields a stable result when integrating downward to lower densities to obtain $\varepsilon(n)$. We then use the thermodynamic identity
    \begin{equation}
        \label{eq:thermoidentity}
        \mu(n) = \frac{\varepsilon(n) + P(n)}{n},
    \end{equation}
and obtain the speed of sound using
\begin{equation}
    \label{eq:cs2mudpdn}
    c_{s}^{2}(n) = \frac{1}{\mu} \frac{\partial P}{\partial n}.
\end{equation}
The undramatic nature of the transition from \ChiEFT\ to pQCD is reflected in our results for the speed of sound shown in Fig.~\ref{fig:informed_priors}(b) and~(d). 
Training the mixed model with data from lower densities in pQCD [case (1)] does result in a more peaked speed of sound in comparison with training only at $n \geq 40 n_0$ [case (2)]. This is related to the GP's tendency to smoothly join the \ChiEFT\ and pQCD curves. Nevertheless, each curve does approach the conformal limit $c_s^2=1/3$ from below in the region where pQCD is valid, also as expected given the pQCD results of Appendix~\ref{ap:cs2_mu} for the consistent $c_s^2$ calculation to $\mathcal{O}(\alpha_s^2)$.

%%%%%%%%%%%%%%%%%%%%%%%%%%%%%%%%%%%%%%%%%%%%%%%%%%%%%%%%%%%%%%%

\section{The perils of pointwise mixing: neglecting correlations necessitates uncontrolled extrapolation} \label{sec:pointwise}

In the previous section, we discussed the results of our mixed model when we employed a GP as the prior on the function space for the EOS and retained the full covariance structure of the input data from the two theories.
We now turn to a pointwise approach, where we use as input completely uncorrelated data (only take as input the means and variances at each point) and construct separate posterior pdfs for the EOS at each point in the input space.
This approach to Bayesian model mixing precision weights the different models being mixed~\mbox{\cite{Phillips:2020dmw}}, and was explored in Ref.~\mbox{\cite{Semposki:2022gcp}} within a toy-model context. Its formalism is described in Sec.~{\ref{sec:MMMformalism}}. As we will see in Sec.~{\ref{sec:MMMresults}}, pointwise mixing requires us to extrapolate the input models into the intermediate region where they do not apply. The extrapolation we employ leads to a very stiff, acausal EOS. These results underscore the need for the GP approach of the previous section, and the importance of the inclusion of correlations in density space in our final, mixed model.

\subsection{Formalism} \label{sec:MMMformalism}

Here, we describe the pointwise approach to Bayesian model mixing, outlined in Ref.~\cite{Phillips:2020dmw} and investigated in Ref.~\cite{Semposki:2022gcp}. This approach is implemented in the versatile BAND collaboration BMM software package \texttt{Taweret}~\cite{Ingles:2023nha}. 
As in Sec.~\ref{sec:GPformalism}, we form
\begin{equation}
    \label{eq:predictions_point}
    Y^{(i)}(x) = F(x) + \delta Y^{(i)}(x), \qquad i \in [1, M],
\end{equation}
where $Y^{(i)}(x)$ are each theory's predictions and $\delta Y^{(i)}(x)$ its truncation errors.

In this section, we discard the information on the correlation structure of $\delta Y^{(i)}(x)$ in density and take only the diagonal elements of the theory covariance matrices derived in Secs.~\ref{subsec:ChiEFT} and \ref{subsec:pQCD}, to explore the impact that this simplification of each model's error structure has on the model mixing. We thus have
\begin{equation}
    \delta Y^{(i)}(x) \sim \mathcal{N}[0, \sigma_{i}^{2}(x)],
\end{equation}
where $\sigma_{i}(x)$ is inferred from the order-by-order convergence of each theory as before.

We again collect the set of model predictions at each point in $x$ into an $M$-dimensional vector: $\vec{y}(x) \equiv \{y_{i}(x)\}$, $i=1, \ldots, M$. The covariance matrix for these data is then denoted by $K_{y}(x)$, as before; however, we now take it to be diagonal: $K_{y,ij}(x) \equiv \sigma_{i}^{2}(x) \delta_{ij}$, $i,j=1, \ldots, M$, therefore assuming that the models are uncorrelated with one another. As mentioned in Sec.~\ref{sec:GPformalism}, this assumption could be relaxed, without modifying any of the equations below, by extending the formalism to pointwise inter-model correlations. But, regardless of that, this approach assumes separate, independent, $M \times M$ matrices $K_y$ at each point $x$ for which we want to perform model mixing. 

The following process is then carried out for each point $x$ in the input space. 
We use Bayes' theorem to write the posterior probability distribution for the underlying theory $F$ at the point $x$:
\begin{align}
\pr(f(x) \given \vec{y}(x), K_{y}(x)) \propto &\pr(\vec{y}(x) \given f(x), K_{y}(x))  \nonumber \\ &\times \pr(f(x)).
\end{align}
As in the previous section, we need to define a prior for the underlying theory; in this situation, we choose
\begin{equation}
    \pr(f(x)) = \mathcal{N}[0, \sigma_{f}^{2}(x)],
\end{equation}
where $\sigma_{f}^{2}(x)$ represents (pointwise) information we have about this theory in the chosen input space. For our present case, we will take this prior to the uninformative limit, $\sigma_{f}^{-2}(x) \rightarrow 0$.

After completing the square, the log posterior for each $f(x)$ is then, up to constant terms,
\begin{align}
    \log &\pr(f(x) \given \vec{y}(x), K_{y}(x)) \propto \nonumber \\
    &-\frac{1}{2} [f(x) - \mu(x)]^T \Sigma^{-1}(x) [f(x) - \mu(x)],
\end{align}
and thus $F(x)$ is Gaussian distributed,
\begin{equation}    
    \pr(f(x)\given\vec{y}(x), K_{y}(x), \sigma_f^{2}(x))=\normal[\mu(x), \Sigma(x)] , \\
\end{equation}
where
\begin{align} \label{eq:pointwise_mu_sigma}
\mu(x) &\equiv \Sigma(x) B^{T} K_{y}^{-1}(x) \vec{y}(x), \nonumber \\
    \Sigma(x) &\equiv \left(\sigma_f^{-2}(x) + B^{T}K_{y}^{-1}(x)B\right)^{-1},
\end{align}
and $B = \vec{1}$ sums the elements of the object it multiplies across the space of $M$ models. 

We have used two (three including the prior) Gaussian random variables that describe the input theories at the point $x$ to infer the mean and variance of the random variable $F(x)$. The combination produces a Gaussian mixed model result, as in the previous section.
Converting the results \eqref{eq:pointwise_mu_sigma} into forms where the sum over model space is explicit we find, in the uninformative limit $\sigma_{f}^{-2}(x) \rightarrow 0$,
    \begin{equation}        
        \label{eq:precisionweighting}
        \mu(x) = \Sigma(x) \sum_{i=1}^{M} \frac{1}{\sigma_{i}^{2}(x)}y_i(x),
        \qquad \Sigma^{-1}(x) \equiv \sum_{i=1}^{M}\frac{1}{\sigma_{i}^{2}(x)}.
    \end{equation}
These formulae are exactly those that govern the estimation of a common mean from $M$ Gaussian-distributed measurements of different precision. The best estimate for the mean is the precision-weighted combination of the measurements, and the variance of that estimate is given by the harmonic mean of the individual measurement variances. In our case this 
combination of the individual model means and variances is done at each point in the input space of the problem, as previously mentioned. This allows the model with the highest precision in a given region to dominate the mixed-model in that region (see Eq.~\eqref{eq:precisionweighting}).

However, the local quality of the mixing means that we need each model being mixed to have calculated values across the entire input space for both $y(x)$ and $\sigma^{2}(x)$. Due to this requirement, in our application, we need to extend the \ChiEFT\ results beyond the density regime where they are typically computed and where \ChiEFT\ with pion-nucleon degrees of freedom is an efficient expansion for nuclear forces in medium ($n \lesssim 2n_0$). 
Although in principle these results are needed up to $100 n_0$, in practice the results in \ChiEFT\ become irrelevant around $(5-6)n_0$, since the estimated pQCD error bars from the present truncation model for $P(n)$ become much smaller than the \ChiEFT\ ones, rendering the \ChiEFT\ weights negligible in the mixture at and beyond that density. Nevertheless, this does mean we need a robust extension of the EOS from where our $\chi$EFT results stop ($\approx 2n_{0}$).

\begin{figure}
    \centering
    \includegraphics[width=\columnwidth]{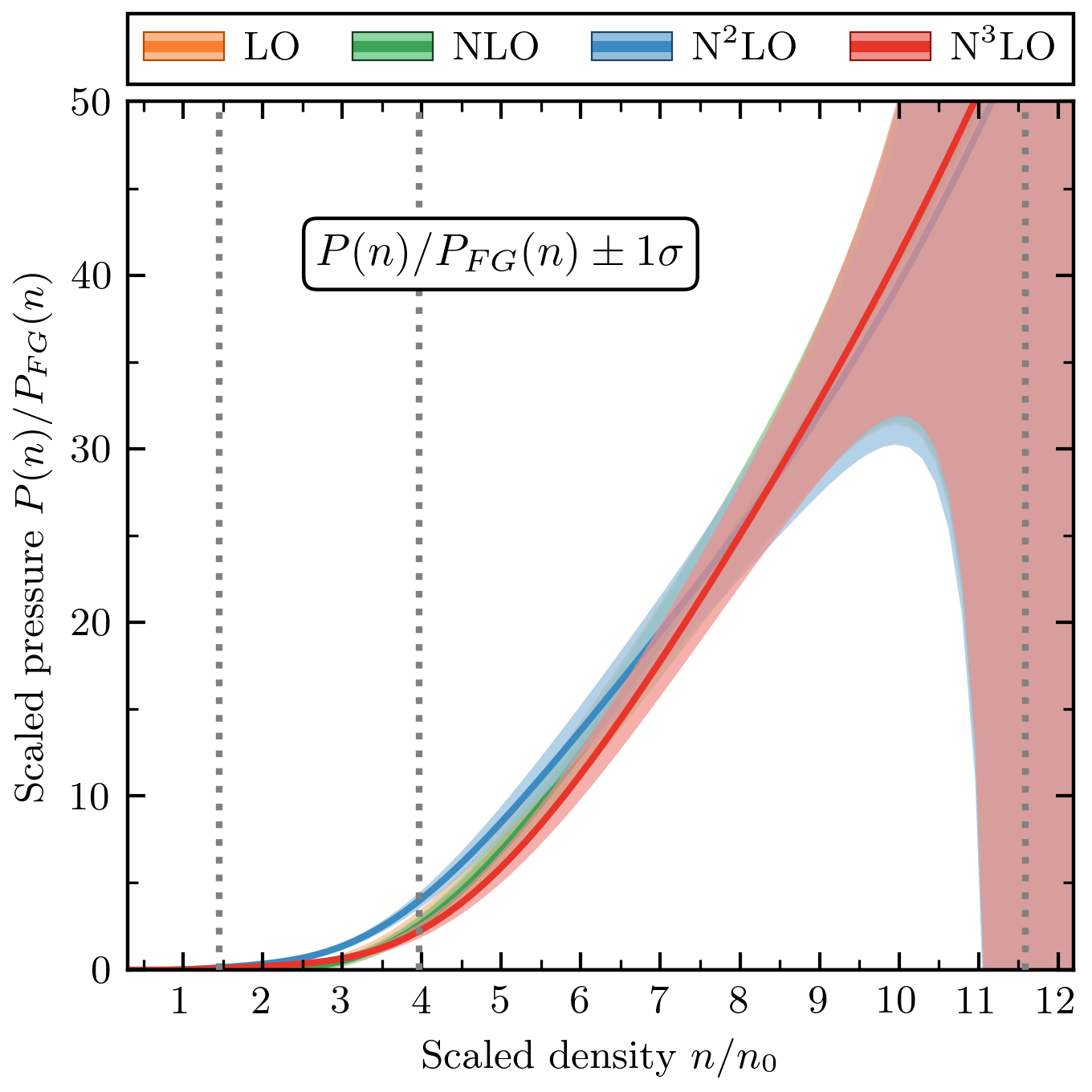}
    \caption{$\chi$EFT pressure scaled by the Fermi gas pressure and extended in the density past the highest density computed in Refs.~\cite{Drischler:2017wtt,Leonhardt:2019fua}, $n \approx 2 n_0$, using the PAL EOS~\eqref{eq:PAL}, which we use here up to $100n_{0}$ (not shown). The dotted vertical lines indicate, from left to right, the densities at which the EFT expansion parameter~\eqref{eq:Q_def} is Q=\{0.5, 0.7, 1.0\}, respectively. Results at lower EFT orders are obscured by those at higher orders in this figure, but we only employ the full result at N$^3$LO and hence focus here on that curve.}
    \label{fig:pressure_scaled_trunc}
\end{figure}

To extrapolate the \ChiEFT\ GP mean function as far as is needed to generate the result shown in Fig.~\ref{fig:pressure_scaled_trunc}, we employ the PAL equation of state \cite{Prakash:1988md}, described in terms of energy per particle as\footnote{To avoid confusion with other quantities in this paper, we divert in Eq.~\eqref{eq:PAL} from the conventional notation of Ref.~\cite{Prakash:1988md}, and use $\beta$ and $\beta'$ in place of $B$ and $B'$, as well as $\zeta$ in place of $\sigma$.}
\begin{align}
    \label{eq:PAL}
    \frac{E(n)}{A} = \frac{3}{5} &E_{F,0} u^{2/3} + \frac{1}{2} A u + \frac{\beta u^{\zeta}}{1 + \beta'u^{\zeta - 1}} \nonumber \\
    &+ 3 \sum_{i = 1,2} C_{i} \left( \frac{\Lambda_{i}}{k_{F,0}} \right)^{3} \left[ \frac{k_{F}}{\Lambda_{i}} - \tan^{-1}\frac{k_{F}}{\Lambda_{i}} \right],
\end{align}
where $u=n/n_0$ is the scaled density and $E_{f,0}=k_{f,0}^{2}/(2m_{N})\approx 37.5 \MeV$ is the Fermi energy at the saturation density. We set the compression modulus $K_{0}=260 \MeV$ and $\beta'=0$, as well as $E(n_0)/A=-16 \MeV$. The corresponding model parameters are $A=-47.84 \MeV$, $\beta = 31.01\MeV$, and $\zeta=1.5$. The finite range parameters are $\Lambda_{1}=1.5 k_{F,0}$, and $\Lambda_{2}=3.0 k_{F,0}$, as in Ref.~\cite{Prakash:1988md}. We also use $C_{1}=-84.84\MeV$ and $C_{2}=23.0\MeV$ (for more details, see Ref.~\cite{Prakash:1988md}). This EOS provides a sufficiently realistic extrapolation into the intermediate region between $\chi$EFT and pQCD for the purposes of this mixing approach. However, we stress that a more realistic intermediate-density parametrization of the nuclear EOS could be used here instead.

The PAL EOS provides the result for the mean value $y(x)$ above $2n_{0}$; to obtain a reasonable description of the truncation error $\sigma(x)$ there, we let the BUQEYE truncation error model continue the calculation of the truncation error on $y(x)$ until the location in $k_{F}$ where the expansion parameter is $Q=1$. After this point, we construct the truncation error such that the PAL EOS has no weight in the mixed model. It is unclear whether the BUQEYE error model describes the truncation error of the \ChiEFT\ calculation beyond $n=2 n_0$, but some extrapolation of the uncertainty bands is necessary if pointwise mixing is to be carried out across the input space.

Even if the PAL EOS, with a continuation of $\chi$EFT error bars, encompasses the QCD result for symmetric nuclear matter between $2 n_0$ and $5 n_0$, pointwise mixing also requires that the pQCD error model be valid down to densities $\approx 2 n_0$. But QCD is no longer perturbative there. While our pQCD error bands can be formally calculated down to the density where $\alpha_s$ diverges---something that actually only occurs at $n< n_{0}$ for our canonical choice $X=1$---they do not account for non-perturbative effects such as renormalons~\mbox{\cite{Beneke:1998ui,Kronfeld:2024qao}} and pairing~\mbox{\cite{Kurkela:2024xfh}}. These are expected to significantly affect QCD thermodynamics for densities below $20 n_0$. There is therefore a significant domain in density for which the outcome of pointwise mixing depends on uncontrolled extrapolations of \mbox{\ChiEFT} and of pQCD uncertainties into regions outside these approaches' region of validity.

\subsection{Implementation and results} \label{sec:MMMresults}

\begin{figure*}[tbh]
    \centering
    \includegraphics[width=\textwidth]{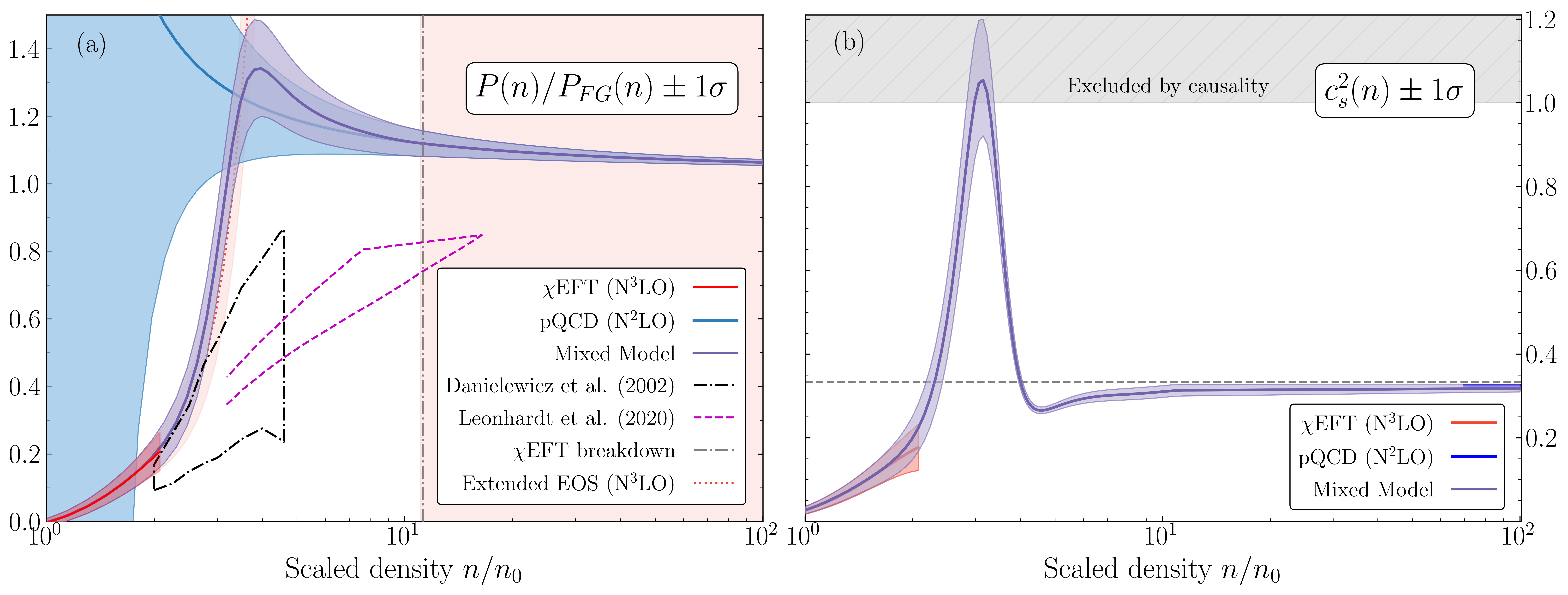}
    \caption{(a) Mixed model results from combining the extended EOS from $\chi$EFT and PAL~\eqref{eq:PAL} with pQCD results for scaled pressure $P(n)/P_{FG}(n)$ using the pointwise model mixing method (Sec.~\ref{sec:pointwise}). Here we use the expansion parameter $Q(\Lambdabar) = N_{f}\alpha_{s}(\Lambdabar)/\pi$. We overlay the results we obtain (purple band) with those of heavy-ion collision data in symmetric nuclear matter (black dash-dot contour) \cite{Danielewicz:2002pu}, and results from the functional renormalization group EOS (FRG) (pink dashed contour) \cite{Leonhardt:2019fua}. We note that the large light red region of the plot (at $n \gtrsim 10 n_0$)  corresponds to the truncation error of the extended EOS past the point where $Q=1$. Because of the sharp increase in the EOS in that density region, the mean curve is no longer visible on the plot, only this portion of the truncation error. (b) The speed of sound squared, $c_{s}^{2}(n)$, for each model. A sharp peak occurs in the region $n \approx 2-5 n_{0}$. The speed of sound squared then drops dramatically and approaches the conformal limit of 1/3 (grey dashed line) from below in the pQCD density region. Due to the peak above unity in $c_{s}^{2}(n)$, this EOS is acausal.} 
    \label{fig:subplot_pointwise_model}
\end{figure*}

We apply our pointwise Bayesian model mixing formalism that combines the two Gaussian distributions for pQCD and \ChiEFT\ to the quantity $P(n)/P_{FG}(n)$, as before. 
Our results are shown in Fig.~\ref{fig:subplot_pointwise_model}.
In Fig.~\ref{fig:subplot_pointwise_model}(a), we plot the scaled pressure, $P(n)/P_{FG}(n)$, in red, blue, and purple with $1\sigma$ uncertainties for $\chi$EFT, pQCD, and the mixed model, respectively. The mixed model follows the $\chi$EFT result until approximately $2n_{0}$, where the theory begins to break down and the truncation error from missing higher-order contributions grows rapidly with density. At this point, because this is a precision-weighted theory at each point in $n$, a mixture of both $\chi$EFT and pQCD is favoured, since the uncertainties of both theories are similar in size. However, past $(3-4)n_{0}$, the uncertainty band of the pQCD EOS decreases to a smaller width than that of $\chi$EFT, and the mixed model begins to more heavily favour pQCD until, at approximately $5n_{0}$, pQCD entirely dominates the mixed model. This result is aligned with our expectations for the size of the individual theories' uncertainty bands. Also plotted in Fig.~\ref{fig:subplot_pointwise_model}(a) are the experimental constraints from heavy-ion collision (HIC) data~\cite{Danielewicz:2002pu}, and the theoretical result from FRG EOS~\cite{Leonhardt:2019fua}, both of which are other information that we have not included in our mixing. The mixed model does not pass through the FRG contour at all, but does fall within the HIC contour until $\approx 3n_{0}$.
    
Figure~\ref{fig:subplot_pointwise_model}(b) shows our result for the speed of sound squared, $c_{s}^{2}(n)$, calculated for $\chi$EFT, pQCD, and the mixed model. This was done by following the procedure described by Eqs.~(\ref{eq:energydensityintegral})--(\ref{eq:cs2mudpdn}) to calculate $c_{s}^{2}(n)$ for each curve. The only difference to the procedure employed in the previous section is that when we compute $c_s^2(n)$ here we employ the result for the energy density at $0.05 \fm^{-3}$ from \ChiEFT\ to set the constant of integration. 

The result of Eq.~\eqref{eq:cs2mudpdn} follows that of $\chi$EFT until just under $2n_{0}$, or about where the mixed model begins to be influenced by the pQCD pressure curve (see Fig.~\ref{fig:subplot_pointwise_model}(a)). The mixed model speed of sound then spikes at $\approx (3-4)n_{0}$, where the pressure curve begins to be dominated by the pQCD EOS. After this point in density, the speed of sound squared falls back down to below the conformal limit of 1/3, and steadily approaches this limit from below until the end of our calculation at $100n_{0}$. This behaviour is anticipated in this density region from our calculation of the sound speed squared from the pQCD EOS alone at N$^2$LO (see Fig.~\ref{fig:pqcd_cs2n.pdf}), which is also shown in Fig.~\ref{fig:subplot_pointwise_model}(b) from $70n_{0}$ onward. The mixed model result is within appropriate error bars of the individual theories, but becomes superluminal at $(3-4)n_{0}$, a result of the rapidly increasing mixed $P(n)/P_{FG}(n)$ curve. This behaviour is not present when we use the GP approach of the previous section. To some extent it is a consequence of the uncontrolled extrapolation of the \ChiEFT\ into a domain where it is not valid, and the use of the pQCD $P(n)$ curve at densities where non-perturbative effects surely  have a significant effect. That is, the pointwise result for $P(n)$ is simply not reliable for $2 n_0 < n < 20 n_0$ because the input to the result in that region is not a controlled approximation to QCD. 

As discussed above, an additional negative consequence of using this pointwise mixing approach is that information on the correlation structure in each model is discarded: we only use the variances of each model at each individual point. This also means that pointwise mixing produces only a diagonal covariance matrix for the mixed model. The mean function happens to be smooth because the data it is fit to is smooth, but this is not guaranteed with this formalism because our prior does not induce any smoothness properties across points. Furthermore, we cannot sample from a probability distribution for curves in the mixed model, because we only have a point-by-point probabilistic representation of it. 

In Ref.~\cite{Semposki:2022gcp}, some of us argued that a prior stating the true function is smooth in the region between the two limiting theories can be included in the analysis by adding a Gaussian process to the mixed model. However, the formalism laid out above, and used in Ref.~\cite{Semposki:2022gcp}, assumes the models being used for the multivariate model mixing are independent, i.e., uncorrelated. This is in fact not the case for the GP constructed in Ref.~\cite{Semposki:2022gcp}, since it was conditioned on training data from the two models describing the low-density and high-density regions of the input space. A GP trained in this way is at least locally 
correlated with the models it is trained on. It is therefore a mistake to mix it with the other two models in a way that does not account for those correlations. Failing to do so effectively double counts information from the region(s) where the GP is trained. 

Since there is no density where both \mbox{\ChiEFT\ } and pQCD converge, the pointwise combination of theories presented in this section necessitates extrapolation of both \mbox{\ChiEFT\ } and pQCD into a region where both break down. With such input, the pointwise mixing approach yields an acausal EOS. The correlated multivariate model mixing approach presented in Sec.~{\ref{sec:curvewise}} is
a conceptually simpler way to include the prior of a smooth connection between the two models in the analysis, and allows us to include more information about the physics in the intermediate region through the use of the physics-informed priors on the mixed-model hyperparameters. The correlated approach also does not double count information from the input models, and makes use of their full covariance matrices, allowing us to take into account the intra-model correlation structure, making this approach the superior method in this application.

%%%%%%%%%%%%%%%%%%%%%%%%%%%%%%%%%%%%%%%%%%%%%%%%%%%%%%%%%%%%%%%

\section{Summary and outlook} \label{sec:summary}

 Recent developments have enabled representations of the EOS of dense QCD matter in the two complementary regimes of ultra-high density and densities up to two times the nuclear saturation density as Gaussian probability distributions. Both calculations---the former from perturbative QCD (pQCD) and the latter from chiral effective field theory (\ChiEFT)---come with correlated theory uncertainties across densities and are derived from the order-by-order convergence pattern of the perturbative series. 
 In this work, we explored a correlated BMM approach
 in which these two (presumably independent) Gaussian random variables can be used to infer what we take to be a common mean, so producing a unified EOS for QCD matter, with quantified uncertainties, for densities all the way from nuclear saturation to asymptotically free quarks (see Fig.~\ref{fig:informed_priors}). 
 Our proof-of-principle calculations carried out this BMM for symmetric, two-flavor strongly interacting matter. 

The two theories used in the mixing were:
\begin{itemize}
\item The \ChiEFT\ pressure at N$^3$LO as a function of number density, including a correlated EFT uncertainty in the density $n$, as computed in Refs.~\cite{Drischler:2017wtt,Drischler:2020hwi,Drischler:2020yad}.

\item The N$^2$LO pQCD pressure as a function of $n$. We convert the N$^{2}$LO results of Ref.~\cite{Gorda:2023mkk} for $P(\mu)$ to $P(n)$ in a way that strictly preserves the perturbative expansion of $P$ using the method of Kohn, Luttinger, and Ward~\cite{Kohn:1960zz, Luttinger:1960ua}. We applied the BUQEYE truncation error model of Ref.~\cite{Melendez:2019izc} to the pQCD EOS to compute the covariance structure associated with N$^3$LO-and-beyond uncertainties in the EOS.
\end{itemize}

The correlated BMM approach to combined inference demonstrated here uses the full structure of both theories' probability distributions to obtain a GP representation for the underlying theory on which both calculations are based. The correlation structure encoded in the mixed model GP kernel through its hyperparameters results in non-local propagation of information from \mbox{\ChiEFT} and pQCD. This allows us to represent each of these two theories using very few training points, and to choose the points representing \mbox{\ChiEFT}/pQCD in that theory's region of validity, thereby avoiding the need to extrapolate either theory outside that domain. In contrast, the pointwise approach of Sec.~{\ref{sec:pointwise}} requires such extrapolations and leads to an acausal EOS.

The correlated approach is based on the following methodological assumption: pQCD and \mbox{\ChiEFT} predictions for the EOS are both statistically related to a common mean function. This assumption results in the uncertainties of the mixed model being (slightly) smaller than the uncertainties of the two input theories even in the regions where one theory dominates the mixing, even when physics-informed priors are employed. This is not really surprising: when two data sets are both used for inference of a common mean the resulting mean is better determined than it would be from either data set alone.

A further and necessary assumption of any Bayesian analysis is the choice of prior on the EOS, which is chosen here to be a stationary GP.
The assumption that a single stationary kernel describes the QCD EOS across all densities means that we can estimate that kernel's length scale and marginal variance from both the pQCD and \mbox{\ChiEFT} training data. The GP that describes matter in the intermediate-density region is then inconceivably well constrained if the MLE estimate of the correlation length from the data sets of \mbox{\ChiEFT} and pQCD is large, as we saw in Fig.{~\ref{fig:joint_subplot_curvewise}}. To remedy this
we assigned hyperpriors to the length scale and marginal variance of the mixed model GP and employed the MAP value instead of the MLE for these quantities. The hyperpriors reflect our knowledge that $\chi$EFT and pQCD uncertainties should be uncorrelated with one another in their respective dominant regions because of the transition from hadron to quark degrees of freedom. Our results for this, our preferred, analysis are shown in Fig.{~\ref{fig:informed_priors}}, for both chosen pQCD cutoffs of $20n_{0}$ and $40n_{0}$.

The details of the uncertainties in the intermediate-density region are affected 
by the prior on the function space in which the correlated mixing takes place, i.e., the choice of GP kernel. In this work, we chose the squared-exponential radial basis function (RBF) kernel to describe our underlying theory uncertainty. This is, in some respects, an extreme assumption since the corresponding function space is stationary and infinitely differentiable. 
We emphasize that the choice of a squared-exponential RBF kernel is not a methodological one; we employed that kernel for demonstrative purposes. Future studies can and should investigate other GP covariance kernels~\mbox{\cite{Mroczek:2023zxo}}.

We emphasize that \emph{QCD need not satisfy any of these assumptions}. What this study shows is that, if the assumptions {\bf are} true, then the EOS at intermediate densities is significantly constrained by the moderate error bars that \mbox{\ChiEFT} and pQCD produce for $P(n)$ at $n < 2 n_0$ and $n > 40 n_0$ respectively (see Fig.{~\ref{fig:informed_priors}}). In fact, the $P(n)$ probability distribution this produces for pQCD training data at $n \geq 40n_{0}$ is consistent with constraints from a Functional Renormalization Group (FRG) calculation~\cite{Leonhardt:2019fua}. The distributions from both pQCD training-data cutoffs are consistent with inference from Heavy-Ion Collisions (HIC)~\cite{Danielewicz:2002pu}. 

The corresponding speed of sound squared, $c_s^2(n)$, is moderately constrained, although whether, and by how much, $c_s^2$ exceeds $1/3$ at $n \approx 10n_0$ depends on the lowest density at which pQCD central values and uncertainties for $P(n)$ are trustworthy. 
Irrespective of the minimum density at which pQCD input is employed, our result for $c_s^2(n)$ approaches $1/3$ smoothly from below as $n$ becomes larger than 40 times the saturation density. This is consistent with the result derived in Appendix~\ref{ap:cs2_mu} that the pQCD speed of sound squared is 1/3 with a calculable correction of $\mathcal{O}(\alpha_s^2)$ whose coefficient is negative because of asymptotic freedom. 

The results produced here could be refined by replacing the N$^2$LO pQCD $P(n)$ by the almost-complete N$^3$LO calculation of Ref.~\cite{Gorda:2023mkk}. Nonzero quark masses~\cite{Kurkela:2009gj,Fraga:2004gz,Gorda:2021gha}, can also be incorporated in the pQCD calculation.
We note, however, that the accuracy of our calculation is not limited by the pQCD input used here, since the error bands we derived for the pQCD $P(n)$ remain narrow down to densities well below those at which non-perturbative effects (e.g., renormalon ambiguities~\cite{Beneke:1998ui,Kronfeld:2024qao} or pairing~\cite{Kurkela:2024xfh}) are expected to markedly affect the pressure. QCD calculations that have reliably quantified uncertainties at intermediate densities where non-perturbative effects play a significant role would be very valuable; rigorous uncertainty quantification for the FRG results~\cite{Leonhardt:2019fua} would be a big step forward in this regard. Any non-perturbative QCD calculation of this character could be straightforwardly added to our multi-model Bayesian inference. 

Potential users of the EOS shown in Fig.~{\ref{fig:informed_priors}} should be aware that this proof-of-principle study is intended to show what can be achieved using the correlated model-mixing framework; our software framework can be easily modified and extended to other use cases through our GitHub repository{~\cite{EOS_BMM_SNM}}. 
The impact that different choices for the GP kernel have on the correlated mixing is an important topic for further study, along with sensitivity analyses that explore the range of acceptable hyperpriors on the GP hyperparameters.

The smooth, squared-exponential RBF kernel also precludes discontinuous phase transitions and other interesting structures of the types discussed in Ref.~\mbox{\cite{Mroczek:2023zxo}}. 
Advanced GP kernels that have limited differentiability and multiple lengthscales can be used to model specific features in the EOS. Their implementation within our framework is a non-trivial extension of the studies carried out here, and will be the subject of a forthcoming paper~\cite{Semposki:2025}. 
The results obtained in the model-mixed EOS with different GP kernels can then be confronted with neutron star observations and data from HIC, which has recently been used for EOS inference~\mbox{\cite{Huth:2021bsp, Yao:2023yda, Komoltsev:2023zor}}. Such data can guide the kernel design and selection.
This process would benefit from, e.g., FRIB400~\cite{FRIB400,LRP:2023}, the proposed 400 MeV/u (uranium) energy upgrade of the Facility for Rare Isotope Beams (FRIB) at Michigan State University, which would provide new experimental constraints on dense matter up to twice the saturation density.

However, the matter described in this work is not neutron-rich matter; extending this study to the case of asymmetric nuclear matter~\cite{Drischler:2020fvz, Kurkela:2009gj, Gorda:2018gpy, Gorda:2021znl, Gorda:2021kme} and specifically
beta-equilibrated, charge-neutral matter (i.e., neutron star matter)~\cite{Lattimer:2000nx, Lattimer:2004pg, Lattimer:2021emm} is a straightforward and important future task. This extension would also provide the opportunity to incorporate the stability constraints discussed in Refs.~\cite{Komoltsev:2021jzg,Komoltsev:2023zor}. The resulting EOSs could be validated against neutron star observations, including NICER and XMM-Newton data (see, e.g., Refs.~\cite{Miller:2021qha, Riley:2021pdl,Salmi:2022cgy,Vinciguerra:2023qxq}), and provide (model-driven) insights into the structure of and maximum sound speed in neutron stars (see, e.g., Refs.~\cite{Drischler:2020fvz,Drischler:2021bup}), thereby complementing the data-driven approaches, e.g., in Refs.~\mbox{\cite{Landry:2018prl,Essick:2020flb,Mroczek:2023zxo}}.
Ultimately, it is also important to extend our methods to mixing across the two-dimensional input space of density and temperature, so providing a unified EOS with quantified uncertainties for numerical simulations of neutron star mergers~\cite{Radice:2018pdn}. 

The inference techniques applied in this paper are in no way specific to $P(n)$ for symmetric two-flavor matter; they apply in any situation where there is a common mean that has been estimated in two independent ways~\cite{BatesGrainger1969}. This work establishes the feasibility and usefulness of this way of combining random variables for dense-matter EOSs, and will provide open-source software to facilitate broad applications of the presented BMM framework.

%%%%%%%%%%%%%%%%%%%%%%%%%%%%%%%%%%%%%%%%%%%%%%%%%%%%%%%%%%%%%%%

\begin{acknowledgments}
We thank Jens Braun, Tyler Gorda, Yoon Gyu Lee, D\'ebora Mroczek, Jaki Noronha-Hostler, and John Yannotty for invaluable insights on several aspects of this work, and Pawel Danielewicz for sharing the HIC constraints on the pressure. We are also grateful to Andrius Burnelis, Andreas Ekstr\"om, Christian Forss\'en, Yuki Fujimoto, Pablo Giuliani, Sudhanva Lalit, Matt Pratola, Frederi Viens, and Christian Weiss for excellent discussions and advice. We acknowledge the hospitality of Chalmers University of Technology (D.R.P. and A.C.S.) and the Facility for Rare Isotope Beams (A.C.S.) during the completion of this work. C.D. thanks the BAND collaboration for their hospitality and encouragement. A.C.S., R.J.F., and C.D. are also all grateful to the Mainz Institute for Theoretical Physics (MITP) of the Cluster of Excellence PRISMA+ (Project ID 390831469), for its hospitality and support. This research was supported by the CSSI program Award OAC-2004601 (BAND collaboration \cite{BAND_Framework}) (A.C.S., R.J.F., D.R.P.), by the U.S. Department of Energy, Office of Science, Nuclear Physics, under Award DE-FG02-93-40756 (A.C.S., D.R.P.), by the FRIB Theory Alliance, under Award DE-SC0013617 (C.D.), by the National Science Foundation Award Nos.\ PHY-2339043 (C.D.) and PHY-2209442 (J.A.M, R.J.F.), by the NUCLEI SciDAC program under award DE-FG02-96ER40963 (R.J.F.), by the Swedish Research Council via a Tage Erlander visiting Professorship, Grant No. 2022-00215 (D.R.P.), and by the US National Science Foundation via Grant PHY-2020275, N3AS PFC (D.R.P.). 
\end{acknowledgments}

%%%%%%%%%%%%%%%%%%%%%%%%%%%%%%%%%%%%%%%%%%%%%%%%%%%%%%%%%%%%%%%

\appendix

\section{Converting pQCD from \texorpdfstring{P($\mu$)}{P(mu)} to P(n)} \label{ap:KLW}

This approach follows the work of Kohn, Luttinger, and Ward \cite{Kohn:1960zz, Luttinger:1960ua}, detailed, e.g., in Ref.~\cite{FETTER71}. First, we define the pressure of perturbative QCD in the manner of Eq.~\eqref{eq:knownorders}, where we choose the expansion parameter $Q(\Lambdabar) = N_{f}\alpha_{s}(\Lambdabar)/\pi$, and the number density as the derivative of the pressure with respect to the quark chemical potential $\mu$, such that
\begin{eqnarray}
    \label{eq:pressnum}
    P(\mu) &=& P_{FG}(\mu) \bigg[c_0 + c_1 Q(\Lambdabar) + c_2(\mu)Q^{2}(\Lambdabar)\bigg], \nonumber \\
    n(\mu) &=& \frac{\partial P(\mu)}{\partial \mu},
\end{eqnarray}
where 
\begin{eqnarray}
    \label{eq:coeffs}
    c_{0} &=& 1, \qquad c_{1} = \frac{a_{1,1}}{N_{f}}, \nonumber \\ 
    c_{2}(\mu) &=& \frac{1}{N_{f}}\bigg(a_{2,1}\ln{\left(\frac{N_{f}\alpha_{s}(\Lambdabar)}{\pi}\right)} \nonumber \\
    &+& a_{2,2}\ln{\left(\frac{\Lambdabar}{2\mu}\right)} + a_{2,3}\bigg).
\end{eqnarray}
Note that the expansion parameter could be chosen differently, and this formalism can easily be altered to incorporate this, with the coefficients altered to account for the difference.

We then define the perturbative expansion of the quark chemical potential as
\begin{equation}
    \label{eq:chempotpert}
    \mu = \mu_{FG} + \mu_{1} + \mu_{2},
\end{equation}
which we only take up to second order due to only having terms in $P(\mu)$ to this order.
We expand the number density of Eq.~\eqref{eq:pressnum} about the point $\mu = \mu_{FG}$, and set
\begin{equation}
    \label{eq:numdens}
    n_{FG}(\mu) = c_{0} \frac{\partial P_{FG}(\mu)}{\partial \mu}\bigg|_{\mu=\mu_{FG}}.
\end{equation}
The number density at the Fermi gas (FG) chemical potential, $n(\mu_{FG})$, is obtained from the first term in this series, and each piece of $\mu$ beyond leading order (LO) is arranged such that the leading order result is unchanged at a specific order in $\alpha_s$. 
Keeping only the terms up to $\alpha_{s}$ and $\alpha_{s}^{2}$, respectively, we thus obtain results for $\mu_{1}$ and $\mu_{2}$
\begin{eqnarray}
    \label{eq:mu1mu2}
    \mu_{1} &=& -\frac{c_{1}Q(\Lambdabar)\frac{\partial P_{FG}(\mu)}{\partial \mu}}{c_{0} \frac{\partial^{2} P_{FG}(\mu)}{\partial \mu^{2}}}\Bigg|_{\mu=\mu_{FG}}, \\
    \mu_{2} &=& -\frac{1}{c_{0} \frac{\partial^{2} P_{FG}(\mu)}{\partial \mu^{2}}} \bigg[ \frac{\mu_{1}^{2}}{2} c_{0} \frac{\partial^{3} P_{FG}(\mu)}{\partial \mu^{3}} \nonumber \\
    &+& \mu_{1}c_{1}Q(\Lambdabar)\frac{\partial^{2}P_{FG}(\mu)}{\partial \mu^{2}} \nonumber \\ 
    &+& c_{1}\frac{\partial Q(\Lambdabar)}{\partial \mu}P_{FG}(\mu) \nonumber \\ 
    &+& \frac{\partial c_{2}(\mu)}{\partial \mu}Q^{2}(\Lambdabar) P_{FG}(\mu) \nonumber \\ 
    &+& c_{2}(\mu)Q^{2}(\Lambdabar)\frac{\partial P_{FG}(\mu)}{\partial \mu} \bigg] \Bigg|_{\mu=\mu_{FG}}.
\end{eqnarray}
    
Now we can recast $\mu_{1}$ and $\mu_{2}$ into expressions containing the number density $n_{FG}(\mu_{FG})$ and simplify to get
\begin{eqnarray}
    \label{eq:chempotwrtn}
    \mu_{1} &=& -\frac{c_{1}Q(\Lambdabar_{FG})\mu_{FG}}{3c_{0}}, \nonumber \\
    \mu_{2} &=& \frac{2c_{1}^{2}}{9c_{0}^{2}}\mu_{FG}Q^{2}(\Lambdabar_{FG}) - \frac{\mu_{FG}}{3c_{0}}c_{2}(\mu_{FG})Q^{2}(\Lambdabar_{FG}) \nonumber \\ 
    &-& \frac{\mu_{FG}^{2}}{12c_{0}}\frac{\partial c_{2}(\mu)}{\partial \mu}Q^{2}(\Lambdabar_{FG})\Bigg|_{\mu=\mu_{FG}} \nonumber \\ 
    &-& \frac{c_{1}}{12c_{0}}\mu_{FG}^{2} \frac{\partial Q(\Lambdabar)}{\partial \mu}\Bigg|_{\mu=\mu_{FG}}.
\end{eqnarray}
We note that $\Lambdabar_{FG} \equiv \Lambdabar(\mu_{FG}) \equiv 2 X \mu_{FG}$, and $\frac{\partial Q(\Lambdabar)}{\partial \mu}$ can be analytically determined at the one-loop level to be \cite{Vermaseren:1997fq}
\begin{equation}
    \label{eq:Qderiv}
    \frac{\partial Q(\Lambdabar)}{\partial \mu} \equiv \frac{N_{f}}{\pi}\frac{\partial \alpha_{s}(\Lambdabar)}{\partial \mu}= -N_{f}\frac{\beta_{0}}{\mu}\frac{\alpha_{s}^{2}(\Lambdabar)}{2\pi^{2}}.
\end{equation}
We can also define
\begin{equation}
    \frac{\partial c_{2}(\mu)}{\partial \mu} = -\frac{1}{N_{f}}\frac{a_{2,1}} {\alpha_s(\Lambdabar)} \frac{\partial \alpha_s(\Lambdabar)}{\partial \mu};
\end{equation}
however, this term is of $\mathcal{O}(\alpha_{s})$, hence inserting for this term in Eq.~\eqref{eq:chempotwrtn} leads to an expression of $\mathcal{O}(\alpha_{s}^{3})$, so we hereafter drop this term from $\mu_{2}$.

We can then expand the pressure in a Taylor series about $\mu=\mu_{FG}$ and keep terms to second order in $\alpha_{s}$, again, to obtain
\begin{eqnarray}
    \label{eq:presstaylorseries}
    P(\mu) &\simeq& c_{0} P_{FG}(\mu_{FG}) \nonumber \\ 
    &+& (\mu_{1} + \mu_{2}) c_{0} \frac{\partial P_{FG}(\mu)}{\partial \mu}\Bigg|_{\mu=\mu_{FG}} \nonumber \\
    &+& c_{1} Q(\Lambdabar_{FG}) P_{FG}(\mu_{FG}) \nonumber \\
    &+& \frac{\mu_{1}^{2}}{2}c_{0}\frac{\partial^{2} P_{FG}(\mu)}{\partial \mu^{2}}\Bigg|_{\mu=\mu_{FG}} \nonumber \\
    &+& \mu_{1}c_{1}Q(\Lambdabar_{FG}) \frac{\partial P_{FG}(\mu)}{\partial \mu}\Bigg|_{\mu=\mu_{FG}} \nonumber \\
    &+& c_{2}(\mu_{FG})Q^{2}(\Lambdabar_{FG})P_{FG}(\mu_{FG}).
\end{eqnarray}
Organising this in terms of powers of $\alpha_{s}$ and inserting $\mu_{1}$ and $\mu_{2}$, and dividing by the FG pressure $P_{FG}(\mu_{FG}(n))$, we can express the scaled pressure with respect to number density $n$ as
\begin{eqnarray}
    \frac{P(n)}{P_{FG}(n)} &=& c_{0} - \frac{1}{3}c_{1}Q(\Lambdabar_{FG}) \nonumber \\
    &+& \frac{2}{9}\frac{c_{1}^{2}}{c_{0}}Q^{2}(\Lambdabar_{FG}) - \frac{1}{3}c_{2}(\mu_{FG})Q^{2}(\Lambdabar_{FG}) \nonumber \\
    &-& \frac{\mu_{FG}}{3}c_{1}\frac{\partial Q(\Lambdabar)}{\partial \mu}\Bigg|_{\mu=\mu_{FG}},
\end{eqnarray}

Now we insert the coefficients from Eq.~\eqref{eq:coeffs} and the expansion parameter as previously mentioned to obtain
\begin{eqnarray}
    \frac{P(n)}{P_{FG}(n)} &=& 1 + \frac{2}{3\pi}\alpha_{s}(\Lambdabar_{FG}) \nonumber \\
    &+& \frac{8}{9\pi^{2}}\alpha_{s}^{2}(\Lambdabar_{FG}) - \frac{N_{f}^{2}}{3\pi^{2}}c_{2}(\mu_{FG})\alpha_{s}^{2}(\Lambdabar_{FG}) \nonumber \\
    &-& \frac{\beta_{0}}{3\pi^{2}}\alpha_{s}^{2}(\Lambdabar_{FG}),
\end{eqnarray}
which is equivalent to Eq.~\eqref{eq:p_nklw}. The first line of this equation contains the LO and NLO terms of the series, and the following lines contain all N$^{2}$LO terms. We have now expressed the pressure in terms of number density, while keeping consistent truncation at second order in the number density and the pressure. Note that this consistent KLW inversion leads to NLO and N$^2$LO coefficients in the pQCD expansion of $P(n)$ that are both smaller, by factors of precisely and roughly $1/3$ respectively, than the coefficients in $P(\mu)$.

\bigskip

\section{Order-by-order speed of sound: \texorpdfstring{P($\mu$)}{P(mu)} and \texorpdfstring{P($n$)}{P(n)} \label{ap:cs2_mu}}

\subsection{Calculating \texorpdfstring{P($\mu$)}{P(mu)}}

We can first define, from QCD, the renormalization group (RG) equation for the running coupling~\cite{ParticleDataGroup:2020ssz},
\begin{equation}
    \label{eq:RGeqn}
    \bar{\Lambda}^{2}\frac{d\alpha_{s}(\bar{\Lambda})}{d \bar{\Lambda}^{2}} = -(b_{0}\alpha_{s}^{2} + b_{1}\alpha_s^{3} + \dots),
\end{equation}
where $b_{0} = \beta_{0}/4\pi$ and $b_{1} = 2\beta_{1}/(4\pi)^{2}$. In this calculation, we will only need to use the one-loop beta function to maintain consistency, so we truncate Eq.~\eqref{eq:RGeqn} after $\mathcal{O}(\alpha_{s}^{2})$.
    
We want to obtain the speed of sound from $P(\mu)$ up to second order in $\alpha_s$, so we will use Eqs.~\eqref{eq:pressurepqcd2023} and~\eqref{eq:soundspeedmu}. Taking the first derivative of Eq.~\eqref{eq:pressurepqcd2023}, we obtain
\begin{eqnarray}
    \label{eq:firstderivPmu}
    \frac{\partial P}{\partial \mu} &=& P_{FG}(\mu) \frac{\partial}{\partial \mu}\left(\frac{P(\mu)}{P_{FG}(\mu)}\right) \nonumber \\ 
    &+& n_{FG}(\mu)\left(\frac{P(\mu)}{P_{FG}(\mu)}\right).
\end{eqnarray}
Then, taking the second derivative of $P(\mu)$, we drop all powers of $\alpha_{s}^{3}$ and above, since we are only interested in the orders up to $\alpha_{s}^{3}$ to maintain thermodynamic consistency. We then series expand in $\alpha_{s}$ and obtain
\begin{equation}
    \label{eq:cs2mu}
    c_{s}^{2}(\mu) \simeq \frac{1}{3} + \mathcal{O}(\alpha_{s}^{2}).
\end{equation}
Hence, no terms at next-to-leading order (NLO) contribute to the sound speed squared with respect to the chemical potential for the pQCD EOS. 

The coefficient of the N$^2$LO term is computed to be 
\begin{equation}
    \label{eq:N2LOcoeff}
    \frac{5}{108\pi^{2}}(-33 + 2N_{f}) \approx -0.136033,
\end{equation}
indicating an approach to the conformal limit from below at the N$^2$LO level for symmetric matter.\footnote{A comprehensive Wolfram Mathematica notebook detailing this calculation is publicly available in the GitHub repository for this project.~\cite{EOS_BMM_SNM}.}

It is interesting to note that, if $\alpha_{s}(\Lambdabar)$ were not a running coupling constant but a simple scalar, the result for the total $c_{s}^{2}(n)$ through N$^2$LO would be 1/3. As such, the speed of sound directly probes the running of $\alpha_{s}(\Lambdabar)$ in pQCD.

\subsection{Calculating \texorpdfstring{P($n$)}{P(n)}}

Due to the relation in Eq.~\eqref{eq:cs2doublelogdvtv}, we can prove in a few simple steps that $c_{s}^{2}(\mu) \equiv c_{s}^{2}(n)$. We do this by using the quark number density and the quark chemical potential $\mu_{q} = \mu_{FG} + \mu_{1} + \mu_{2}$ from Appendix~\ref{ap:KLW}, which implicitly contains the number density. We can write
\begin{widetext}
\begin{equation}
    \label{eq:cs2n_klw}
    c_{s}^{2}(\mu(n)) = \frac{n}{\mu} \frac{\partial \mu}{\partial n}\Bigg|_{n=\bar{n}} \equiv \frac{\bar{n}}{(\mu_{FG} + \mu_{1} + \mu_{2})} \frac{\partial (\mu_{FG} + \mu_{1} + \mu_{2})}{\partial n}\Bigg|_{n=\bar{n}} \equiv \frac{c_{s,FG}^{2}(\bar{n})}{\left(1 + \frac{\mu_{1}}{\mu_{FG}} + \frac{\mu_{2}}{\mu_{FG}}\right)} \left(1 + \frac{\partial \mu_{1}}{\partial \mu_{FG}} + \frac{\partial \mu_{2}}{\partial \mu_{FG}}\right),
\end{equation}
\end{widetext}
where the Fermi gas sound speed squared is given by
\begin{equation}
    c_{s,FG}^{2}(\bar{n}) = \frac{\bar{n}}{\mu_{FG}}\frac{\partial \mu_{FG}}{\partial n}\Bigg|_{n=\bar{n}}.
\end{equation}
Here, $\bar{n}$ is a given quark number density. 

Taking a series expansion of the denominator in Eq.~\eqref{eq:cs2n_klw} and truncating any terms we know will yield $\mathcal{O}(\alpha_{s}^{3})$ or higher immediately, we obtain
\begin{eqnarray}
    \label{eq:series_cs2n}
    c_{s}^{2}(\mu(n)) &=& c_{s,FG}^{2}(\bar{n})\Bigg[1 - \frac{\mu_{1}}{\mu_{FG}} - \frac{\mu_{2}}{\mu_{FG}} + \frac{\mu_{1}^{2}}{\mu_{FG}^{2}} \nonumber \\ 
    &+& \frac{\partial \mu_{1}}{\partial \mu_{FG}} - \frac{\mu_{1}}{\mu_{FG}}\left(\frac{\partial \mu_{1}}{\partial \mu_{FG}}\right) - \frac{\mu_{2}}{\mu_{FG}} \nonumber \\ 
    &+& \left(\frac{\partial \mu_{2}}{\partial \mu_{FG}}\right) + \mathcal{O}(\alpha_{s}^{3})\Bigg].
\end{eqnarray}

From Eqs.~\eqref{eq:chempotwrtn}, we can construct the expressions for all of the terms in Eq.~\eqref{eq:series_cs2n}. After simplification and use of Eq.~\eqref{eq:Qderiv}, these are
\begin{eqnarray}
    \label{eq:simplifiedeqns_cs2_pieces}
    \frac{\mu_{1}}{\mu_{FG}} &=& \frac{2}{3\pi}\alpha_{s}(\Lambdabar), \nonumber \\
    \frac{\mu_{2}}{\mu_{FG}} &=& \frac{8}{9\pi^{2}}\alpha_{s}^{2}(\Lambdabar) - \frac{4}{3\pi^{2}}c_{2}(\mu_{FG})\alpha_{s}^{2}(\Lambdabar) \nonumber \\ 
    &+& \frac{\mu_{FG}}{3\pi}\alpha_{s}'(\Lambdabar), \nonumber \\ 
    \frac{\partial \mu_{1}}{\partial \mu_{FG}} &=& \frac{\mu_{1}}{\mu_{FG}} + \frac{4}{3\pi}\mu_{FG}\alpha_{s}'(\Lambdabar), \nonumber \\
    \frac{\partial \mu_{2}}{\partial \mu_{FG}} &=& \frac{\mu_{2}}{\mu_{FG}} + \frac{\mu_{FG}}{3\pi}\alpha_{s}'(\Lambdabar),
\end{eqnarray}
where $\alpha_{s}'(\Lambdabar) = \partial \alpha_{s}(\Lambdabar)/\partial \Lambdabar$.

Next, we insert these expressions in Eq.~\eqref{eq:series_cs2n}, truncate any terms above $\mathcal{O}(\alpha_{s}^{2})$, achieving the relation
\begin{eqnarray}
    \label{eq:final_cs2n_klw}
    c_{s}^{2}(\mu(n)) &=& c_{s,FG}^{2}(\bar{n}) \left(1 + \frac{5}{3\pi}\mu_{FG}\alpha_{s}'(\Lambdabar)\right), \nonumber \\
    &=& \frac{1}{3}\left(1 - \frac{5}{3\pi}\beta_{0}\alpha_{s}^{2}(\Lambdabar)\right),
\end{eqnarray}
where we have inserted $c_{s,FG}^{2}(\bar{n}) = \frac{1}{3}$ and the derivative $\partial \alpha_{s}(\Lambdabar)/\partial \Lambdabar = -(2/\Lambdabar)\alpha_{s}^{2}(\Lambdabar)$, which can be seen from Eq.~\eqref{eq:RGeqn}. This calculation numerically yields an N$^{2}$LO coefficient of
\begin{equation}
    \label{eq:cs2_N2LOcoeff_n}
    -\frac{5}{108\pi^{2}}(33 - 2N_{f}) \approx -0.136033,
\end{equation}
which exactly matches the result from Eq.~\eqref{eq:N2LOcoeff}, where we calculated $c_{s}^{2}(\mu)$ directly from the pQCD pressure (see Eq.~\ref{eq:pressurepqcd2023}). This argument proves that the KLW inversion yields the correct relation for both $P(n)$ and $c_{s}^{2}(n)$ up to and including N$^{2}$LO, further validating our procedure in Appendix~\ref{ap:KLW}.

%%%%%%%%%%%%%%%%%%%%%%%%%%%%%%%%%%%%%%%%%%%%%%%%%%%%%%%%%%%%%%%%%

%\clearpage
\bibliography{bayesian_refs}

%%%%%%%%%%%%%%%%%%%%%%%%%%%%%%%%%%%%%%%%%%%%%%%%%%%%%%%%%%%%%%%%

\end{document}

%% file: buqeye_macros.tex
\makeatletter
\newcommand\newsubcommand[3]{\newcommand#1{#2\sc@sub{#3}}}
\def\sc@sub#1{\def\sc@thesub{#1}\@ifnextchar_{\sc@mergesubs}{_{\sc@thesub}}}
\def\sc@mergesubs_#1{_{\sc@thesub#1}}

\newcommand\newsupcommand[3]{\newcommand#1{#2\sc@sup{#3}}}
\def\sc@sup#1{\def\sc@thesup{#1}\@ifnextchar^{\sc@mergesups}{^{\sc@thesup}}}
\def\sc@mergesups^#1{^{\sc@thesup#1}}
\makeatother

\DeclareMathAlphabet{\mathbcal}{OMS}{cmsy}{b}{n}

\newcommand{\cbar}{\bar c}

\newcommand{\fm}{\, \text{fm}}

\newcommand{\fmiq}{\, \text{fm}^{-3}}

\newcommand{\MeV}{\, \text{MeV}}
\newcommand{\ordervec}{\vec}

\newcommand{\inputvec}{\mathbf}

\newsubcommand{\ckvec}{\ordervec{c}}{k}

\newsubcommand{\bkvec}{\ordervec{b}}{k}
\newsubcommand{\ckvecset}{\ordervec{\inputvec{c}}}{k}

\newsubcommand{\ckvecapprox}{\mathbf{c}'}{k}
\newsubcommand{\ckvecapproxset}{\mathbf{C}'}{k}

\newsubcommand{\bkvecapprox}{\mathbf{b}'}{k}
\newsubcommand{\bkvecset}{\mathbf{B}}{k}
\newsubcommand{\bkvecapproxset}{\mathbf{B}'}{k}

\newcommand{\genobs}{y}

\newsubcommand{\genobsvec}{\ordervec{\genobs}}{k}
\newsubcommand{\genobsvecset}{\ordervec{\inputvec{\genobs}}}{k}

\newsubcommand{\akvec}{\mathbf{a}}{k}

\newsubcommand{\akvecapprox}{\mathbf{a}'}{k}
\newsubcommand{\akvecset}{\mathbf{A}}{k}
\newsubcommand{\akvecapproxset}{\mathbf{A}'}{k}

{}  % Remove the definition from the Physics package

\DeclareMathOperator{\pr}{pr} % Good, but want to handle sizing and | spacing?
\newcommand{\given}{\,|\,}  % Use for | in \pr

\newcommand{\normal}{\mathcal{N}}

\def\diffd{\mathrm{d}}  % Upright differentials
\DeclareDocumentCommand\differential{ o g d() }{ % Differential 'd'
    \IfNoValueTF{#2}{
        \IfNoValueTF{#3}
            {\diffd\IfNoValueTF{#1}{}{^{#1}}}
            {\mathinner{\diffd\IfNoValueTF{#1}{}{^{#1}}\argopen(#3\argclose)}}
        }
        {\mathinner{\diffd\IfNoValueTF{#1}{}{^{#1}}#2} \IfNoValueTF{#3}{}{(#3)}}
    }
\DeclareDocumentCommand\dd{}{\differential} % Shorthand for \differential

\newcommand{\pathd}{\mathcal{D}}  % differential symbol for path integrals

\DeclareDocumentCommand\pathdifferential{ o g d() }{ % Path 'D'
    \IfNoValueTF{#2}{
        \IfNoValueTF{#3}
            {\pathd\IfNoValueTF{#1}{}{^{#1}}}
            {\mathinner{\pathd\IfNoValueTF{#1}{}{^{#1}}\argopen(#3\argclose)}}
        }
        {\mathinner{\pathd\IfNoValueTF{#1}{}{^{#1}}#2} \IfNoValueTF{#3}{}{(#3)}}
    }